\documentclass[12pt,preprint]{aastex}

\usepackage{amsmath,amssymb} 
\usepackage{subfigure} 

\shorttitle{Active Layers of Disks}
\shortauthors{Bai and Goodman}

\newcommand{\au}{\,\textsc{au}} 
\newcommand{\dash}{\mbox{-}} 
\newcommand{\K}{{\rm K}} 
\newcommand{\kB}{k_{\textsc{b}}} 
\newcommand{\Htwo}{{\rm H_2}}
\newcommand{\water}{\,{\rm H_2O}\,}
\newcommand{\Siga}{\Sigma_{\rm a}}

\begin{document}

\title{Heat and Dust in Active Layers of Protostellar Disks}


\author{Xue-Ning Bai and Jeremy Goodman}
\affil{Princeton University Observatory, Princeton, NJ 08544}

\begin{abstract}
  Requirements for magnetic coupling and accretion in the active layer
  of a protostellar disk are re-examined, and some implications for
  thermal emission from the layer are discussed.  The ionization and
  electrical conductivity are calculated following the general scheme
  of Ilgner and Nelson but with an updated UMIST database of chemical
  reactions and some improvements in the grain physics, and for the
  minimum-mass solar nebula rather than an alpha disk. The new limits
  on grain abundance are slightly more severe than theirs. Even for
  optimally sized grains, the layer should be at least marginally
  optically thin to its own thermal radiation, so that narrow, highly
  saturated emission lines of water and other molecular species would
  be expected if accretion is driven by turbulence and standard rates
  of ionization prevail. If the grain size distribution extends
  broadly from well below a micron to a millimeter or more, as
  suggested by observations, then the layer may be so optically thin
  that its cooling is dominated by molecular emission.  Even under
  such conditions, it is difficult to have active layers of more than
  $10{\rm g\,cm^{-2}}$ near $1\au$ unless dust is entirely eliminated
  or greatly enhanced ionization rates are assumed.
  Equipartition-strength magnetic fields are then required in these
  regions of the disk if observed accretion rates are driven by
  magnetorotational turbulence. Wind-driven accretion might allow
  weaker fields and less massive active layers but would not heat the
  layer as much as turbulence and therefore might not produce emission
  lines.

\end{abstract}


\keywords{accretion disks---magnetic fields---molecular
  processes---stars:pre-main sequence}


\goodbreak
\section{Introduction}
\label{sec:intro}

Observations of classical T Tauri stars indicate accretion rates
$\dot{M}\sim10^{-8\pm1}M_{\odot}$yr$^{-1}$ for $10^{6-7}{\rm
  yr}$.\citep{Hartmann_etal98}. The torques driving the accretion are
probably magnetic, whether exerted on the face of the disk by a
magnetized wind \citep{Blandford_Payne82}, or across the thickness
of the disk by magnetorotational (MRI) turbulence
(\citealp{BH91,BH98}). At the low temperatures characteristic of
most of the mass of a protostellar disk, the ionization fraction
would be too low for good coupling to a magnetic field were it not
for nonthermal ionization processes such as cosmic rays, X-rays, and
radioactivity. Since the sources of the first two lie outside the
disk, the surface layers of the disk are expected to be more
strongly ionized than the midplane. It is possible that accretion
occurs mainly in the surface layers, while the rest of the disk
column forms a magnetically decoupled ``dead zone'' in which mass
may accumulate \citep{Gammie96}.  Dead zones may be favorable
environments for the formation of planets, or for the survival of
planets once formed (e.g. \citealp{Terquem08,Ida_Lin08}).

It has been recognized at least since Gammie's work that small dust
grains, through their influence on the ionization balance, are
crucial for MRI turbulence and for magnetic coupling more generally.
A number of authors have investigated the thickness of the active
layer based on different disk models, ionization sources, chemical
reaction networks, and grain properties. \citet{Glassgold_etal97}
and \citet{Igea_Glassgold99} pointed out the importance of stellar
X-rays to the active layer. \citet{Sano_etal00} considered the
minimum mass solar nebula (MMSN, \citealp{Hayashi81}) with a
chemical network including dust grains and found dead zones
extending to $\sim15\au$. \citet{Fromang_etal02} showed that the
conductivity is sensitive to the abundance of ``metals'' (Mg, Fe,
etc.) in the gas phase, eliminating the dead zone entirely from some
of their grain-free models, which were $\alpha$~disks with lower
column density than the MMSN. \citet{Semenov_etal04} adopted the
``fiducial'' $\alpha$~disk model by \citet{D'Alessio_etal99}, which
also has lower surface density at $1\au$ than the MMSN, and a
chemical network based on the UMIST~95 database
\citep{Millar_etal97} supplemented by reactions on grain surfaces
from \citet{Hasegawa_etal92}; they found a marginally dead zone
somewhat smaller than that of \citet{Sano_etal00}. More recently, in
a similar study based on an $\alpha$-disk model and a yet more
extensive network, \citet{IlgnerNelson06a}
concluded that adding sub-micron sized grains with concentration of
$x_{\rm gr}=10^{-12}$ per ${\rm H_2}$ molecule efficiently depletes
metal atoms and dramatically reduces the extent of the active layer.

Dust grains in protostellar disks have observable signatures in the
spectral energy distribution (SED) from infrared to millimeter
wavebands. Small grains in particular near the disk surface are
superheated by absorbtion of visible light from the star; this
promotes flaring (concavity) of the disk surface and may contribute
to the relatively slow decline of the SED with increasing wavelength
\citep{Chiang_Goldreich97,Chiang_Goldreich99}. More refined models,
however, show that the dust size distribution may not be well
constrained solely from SEDs \citep{Chiang_etal01}.
\citet{D'Alessio_etal06} further studied comprehensively the effect
of grain growth and settling on the SED of the protoplanetary disks;
comparing their theoretical SED with recent mid- and near-IR
observations (e.g., \citealp{Hartmann_etal05}), they found that
grains in the disk upper layer should be depleted by at least a
factor of 10 relative to interstellar dust-to-gas ratios, and
perhaps by $10^2-10^3$.

Little attention has been given to the implications of the
conductivity constraints for the emissions by the dust and gas. This
is the focus of the present paper. We recalculate, in yet greater
detail, the influence of grains on the ionization balance. In view
of the ambiguities in measuring small grains from the SEDs and the
difficulty of improving on previous work on that problem, however,
we chose to concentrate on an indirect signature of dust depletion:
enhanced molecular emission from the active layer. The fundamental
CO emission line has been frequently observed in CTTSs, with the
likely origin of the protostellar disks within 1-2 AU (e.g.,
\citealp{Najita_etal03}). Recently, discoveries of organic and water
molecules have been reported, and these molecules are likely to
originate from the inner region ($<5\au$) of young protostellar
disks \citep{Carr_Najita08,Salyk_etal08,Mandell_etal08}. These
molecular emissions are consistent with hot, optically thin gas.
Based on the derived temperature and column density, the emission
lines may originate from a UV-heated layer above the disk surface.
However, we argue from the conductivity constraints that dust must
be sufficiently depleted from the active layer so that it is at
least marginally optically thin, allowing molecular emission lines
to stand out. If the dust size distribution extends to grains as
small as $10^{-2}\micron$, then the heat of accretion may be
expressed primarily in these lines rather than in the dust
continuum.

The organization of the paper is as follows. In
\S\ref{sec:magnetic}, we review constraints on the minimum
ionization fraction and magnetic field strength required to drive
accretion at observed rates by winds or by MRI turbulence. In
\S\S3-4, we calculate the thickness of the active layer. This is an
extension of the work by IN06a, with the following differences and
additions. Firstly, we use the MMSN for the density and temperature
profile of the disk, rather than an $\alpha$ model. The MMSN is
simply defined and has been used extensively as a standard model for
discussion of protostellar disks and planet formation. It has an
empirical basis in the observed properties of our own solar system
and is not inconsistent with inferences of T~Tauri disks by, e.g.,
submillimeter emission. Steady $\alpha$ disks are have a weaker
observational basis. They come in many forms, depending on
assumptions about accretion and cooling, and they are often not
self-consistent where active layers exist. The particular $\alpha$
model used by IN06a is convex rather than flared like the MMSN.
Secondly, we adopt the latest version of the UMIST database
\citep{Woodall_etal07} for the chemical reactions. Changes in the
reaction rates have perceptible consequences for the ionization
fraction. Thirdly, we provide more detailed analysis of the
dependence of conductivity on the grain properties, especially their
size distribution. In \S5, we discuss the conditions under which
molecular lines in the active layer of protostellar disks may stand
out above the dust continuum. For illustrative purposes, we study
the water molecule only; it is abundant, and its rich spectrum makes
it a good coolant. Using the emission line and energy level list
given by \citet{Barber_etal06} (hereafter BT), we calculate the
molecular line emissions from an isothermal slab in local
thermodynamic equilibrium (LTE) to mimic the disk active layer.
Various line broadening effects are considered in our calculation.
We derive the conditions on the dust abundance and size distribution
in which water lines dominate the cooling of the layer. Limitations
of our study and some open questions are discussed in \S6, and a
summary is given in \S7.

\goodbreak
\section{Requirements for Magnetically-Driven Accretion}
\label{sec:magnetic}

In this section, we discuss theoretical constraints on the
ionization fraction ($x_e$), magnetic field strength ($B_z$ or
$B_{\rm rms}$), and surface mass density ($\Siga$) of the active
layer required to explain magnetically driven accretion rates of
order $10^{-7}\,\dot M_{-7}\,{\rm M_\sun\,yr^{-1}}$, where $\dot M_{-7}
=\dot M/(10^{-7}{\,\rm M_\odot\,yr^{-1}})$. First we
consider a magnetized wind, and then MRI turbulence.  In the latter
case, we emphasize the conditions needed to sustain accretion at the
above rate, which are more stringent than those required for linear
instability.  The minimal field strength ($B$) is about one order of
magnitude larger for MRI turbulence than for wind-driven accretion
at the same $\dot M$.  But the degree of ionization is similar.

\subsection[]{Disk Model}\label{ssec:disk}

We adopt the minimum-mass solar nebular (MMSN) model with stellar
mass $M_*=1M_{\bigodot}$ \citep{Hayashi81,Hayashi_etal85}. The disk
is vertically isothermal with radial temperature profile
\begin{equation}
T=T_0r_{\au}^{-1/2}\,\qquad T_0=280\K,
\end{equation}
surface mass density
\begin{equation}
\Sigma=\Sigma_0r_{\au}^{-3/2}\,\qquad \Sigma_0=1700\rm{g\ cm}^{-2}\,,
\end{equation}
sound speed
\begin{equation}
  \label{eq:soundspeed}
c_s=\left(\frac{kT}{\mu m_H}\right)^{1/2}=0.99\,r_{\au}^{-1/4}{\rm km\,s^{-1}}\,,
\qquad \mu\approx 2.34,
\end{equation}
volume density
\begin{equation}\label{eq:rho}
\rho(r,z)=\rho_0(r)\exp(-z^2/2h^2)\,,\qquad
\rho_0(r)=\frac{\Sigma}{h\sqrt{2\pi}}
=1.4\times10^{-9}r_{\au}^ {-11/4}{\rm g\,cm^{-3}}\,,
\end{equation}
and scale height
\begin{equation}
h=c_s/\Omega_K=0.03r_{\au}^{5/4}{\rm{AU}}\ ,
\end{equation}
where $r_{\au}$ is the disk radius measured in astronomical units
and $\Omega_K=(GM_*/r^3)^{1/2}$. Note that the disk flares, that is,
$h/r$ increases with $r$.

\subsection{Magnetized wind}\label{subsec:wind}
When accretion is wind-driven, angular momentum is extracted via the
mean magnetic torque per unit area $-r\overline{B_z B_\phi}/4\pi$
exerted on the surface the disk.  This must balance the rate of loss
of angular momentum per unit area, which is $-\dot M\Omega/8\pi$
when one considers that the accretion is divided between the two
faces of the disk so that
\begin{equation}
  \label{eq:windbalance}
(\overline{|B_z|})(\overline{|B_\phi|})\ge\dot M\Omega/2r\,.
\end{equation}
We consider $\dot M>0$ for inflow, and hereafter we omit the
overbars and absolute value signs where this will not cause
confusion.  It is also required that $B_r\ge B_z/\sqrt{3}$ in order
that the wind be centrifugally propelled \citep{Blandford_Payne82}.
Minimizing the total magnetic energy subject to these constraints
leads to a minimum total field at the surface of the disk
\citep[e.g.][]{Wardle07},
\begin{equation}
  \label{eq:windfield}
  B\ge\left(\frac{2}{\sqrt{3}}\frac{\dot M\Omega}{r}\right)^{1/2}\approx
 0.31\,r_{\au}^{-5/4}\,\dot M_{-7}^{1/2}\,\mbox{G}\ .
\end{equation}
In this configuration, $B_r=-B_\phi/2$; some wind models require
$|B_r|>|B_\phi|$ (e.g., \citealt{Wardle_Koenigl93}) and therefore a
stronger total field for the same $\dot M$.

The active layer has thickness $h_a\approx \xi c_s/\Omega$, where
the dimensionless factor $\xi$ is such that the volume density at
the base of the active layer is $\rho_a=\Sigma_a/h_a$.  Since the
vertical density profile under isothermal conditions is gaussian
with scale height $c_s/\Omega$, $\xi$ depends logarithmically on
$\Sigma_a/\Sigma$; for example, if $\Sigma_a=10\,{\rm g\,cm^{-2}}$,
then the appropriate value of $\xi\approx0.39$ at $1\,\au$.  The
roughness of our estimates hardly justifies keeping track of this
logarithmic dependence, so we simply take $h_a=0.5c_s/\Omega$ at all
radii hereafter.

In the configuration that achieves the minimum (\ref{eq:windfield}),
the total horizontal component $B_h=(B_r^2+B_\phi^2)^{1/2}$ is
approximately equal to $|B_z|$ at the disk surface.  But in the dead
zone, $B_h\ll B_z$ because the field there is poorly coupled to the
gas and straightens out under magnetic tension.  Thus in the active
layer, $|\partial B_h/\partial z|\gtrsim |B_z|/h_a\gg |\partial
B_z/\partial r|$, so that there will be a horizontal electric
current in the active layer and an associated electric field.  As
described in the review by \cite{Wardle07}, the current and the
electric field are not in the same direction under the likely
conditions prevailing in active layers; the conductivity is
nontrivially tensorial because of the field itself, and because some
charged species are more firmly attached to the field than others.
This is quantified by the Hall parameters
$\beta_j\equiv\omega_{j}t_{j}$ for each charged species $j$, where
$\omega_j=|q_jB|/m_j c$ is the cyclotron frequency, and
$t_{j}=m_jm_n/\mu_j\rho\langle\sigma v\rangle_{jn}$ is the timescale
on which particles of species $j$ lose their momentum in collisions
with the neutrals.  The reduced mass $\mu_j\equiv m_jm_n/(m_j+m_n)$
is essentially $m_e$ for free electrons, and $\mu_j\approx
m_n\approx 2m_p$ for metallic or molecular ions as well as for
charged grains, since all of these are heavier than molecular
hydrogen.  Wardle gives numerical scalings for free electrons and
ions, which we rewrite in terms of $\Sigma_1\equiv\Sigma_a/(10\,{\rm
  g\,cm^{-2}})$, $B_{\rm G}\equiv B/(1\,{\rm G})$, and the standard
properties of the MMSN as
\begin{equation}
  \label{eq:betas}
  \beta_e\approx 87\,\Sigma_1^{-1}B_{\rm G}\,r_{\au}^{3/2}\,,\quad
  \beta_i\approx 0.19\,\Sigma_1^{-1}B_{\rm G}\,r_{\au}^{5/4}\,.
\end{equation}
Charged grains are unmagnetized, $\beta_g\ll 1$. For magnetic fields
comparable to the value (\ref{eq:windfield}), the ions follow the
neutrals (because $\beta_i\ll 1$) but the electrons follow the field
(because $\beta_e\gg1$).  In the rest frame of the neutrals, the
current density $\boldsymbol{J}$ is therefore borne almost entirely
by the electrons, as is the Lorentz force $\boldsymbol{J\times
B}/c$.  So there must be an electric field $\boldsymbol{E}\approx
-\boldsymbol{J\times B}/e n_e c$ in this frame to prevent the
electrons from accelerating, since collisions are ineffective.  Thus
``Ohm's Law'' in the frame of the neutrals becomes
$\sigma_H\boldsymbol{E}\approx\boldsymbol{J\times\hat B}$, where
$\boldsymbol{\hat B}\equiv\boldsymbol{B}/B$ is a unit vector and
$\sigma_H\approx -en_e c/B$ is the Hall conductivity.  The electric
field component parallel to $\boldsymbol{B}$ is much smaller than
the perpendicular component because the ordinary Ohmic conductivity
$\sigma_O\approx e^2 n_e t_e/m_e=-\beta_e\sigma_H$ is much larger
than $\sigma_H$.

The electric field in the inertial reference frame of the star,
where the neutrals have velocity $\boldsymbol{v}$, is
$\boldsymbol{E'}=\boldsymbol{E}- \boldsymbol{v\times B}/c$.  The
azimuthal component $E'_\phi$ must vanish, at least on average,
provided that there is no secular increase in the magnetic flux
threading the disk.  Substituting
$\boldsymbol{E}\approx\sigma_H^{-1}\boldsymbol{J\times\hat B}\approx
(\boldsymbol{B\times\nabla\times B})/4\pi en_e$ into
$E'_\phi=E_\phi+(v_rB_z-v_zB_\phi)/c$, and presuming that $v_z\ll
v_r\ll v_\phi$ within the active layer (though not within the wind),
we find that the steady-state accretion velocity in the layer is
related to the the electron density by
\begin{equation}
  \label{eq:vrdrift}
  v_r\approx -\frac{c}{4\pi e n_e}\frac{\partial B_\phi}{\partial z}\,.
\end{equation}
On the other hand, local angular momentum conservation requires
\begin{equation}
  \label{eq:phiforce}
v_r\frac{d}{dr}(\Omega r^2)=\frac{B_z}{4\pi\rho}\frac{\partial
  B_\phi}{\partial z}
\end{equation}
Eliminating $B_\phi$ and $v_r$ between these two expressions leads to
\begin{equation}
  \label{eq:xeB}
  x_e\equiv\frac{n_e}{n_H}\approx -\frac{c\Omega\rho}{2e B_z n_H}\,,
\end{equation}
where $n_H$ is the density of hydrogen nuclei, so that
$\rho/n_H\approx 1.4 m_p$ at cosmic abundance.  Taking
$B_z^2=-(3/4)B_\phi B_z=3\dot M\Omega/8r$, as is true for the
minimal configuration (\ref{eq:windfield}), we have
\begin{equation}
  \label{eq:xeM}
  x_e\approx 8.3\times 10^{-11}\dot M_{-7}^{-1/2}r_{\au}^{-1/4}\,,
\end{equation}
which is independent of $\Sigma_a$. Note that
$\boldsymbol{B\cdot\Omega}<0$ in order that $x_e>0$ in
eq.~(\ref{eq:xeB}). Strictly steady wind-driven accretion in the
extreme Hall limit ($\beta_i\ll 1\ll\beta_e$) appears to be a
degenerate case. It may be that the active layer for wind-driven
accretion must be in the ambipolar ($1\lesssim\beta_i\ll\beta_e$)
regime, as in the models of \cite{Wardle_Koenigl93}. Still, the
electron fraction~(\ref{eq:xeM}) is interesting as a characteristic
value.

A lower bound on $\Sigma_a$ can be obtained by considering that the
gas pressure at the base of the layer can be no smaller than the
change in magnetic pressure across the layer.  The former is $P_{\rm
gas}=\Sigma_a c_s^2/h_a\approx 2\Sigma_a c_s\Omega$, unless the
field is strong enough to significantly compress the layer (which is
actually a requirement of the models of \citealt{Wardle_Koenigl93}).
The latter is $\Delta P_{\rm mag}=(B_r^2+B_\phi^2)/8\pi$, which
becomes $(5/8)B^2/8\pi$ in the minimum-energy magnetic configuration
(\ref{eq:windfield}).  Hence
\begin{equation}
  \label{eq:Sigma_min_wind}
  \Sigma_a\gtrsim 0.061\dot M_{-7} r_{\au}^{-3/4}\,{\rm g\,cm^{-2}}.
\end{equation}
This is well into the ambipolar regime, because the ion Hall
parameter $\beta_i>1$ for $\Sigma<0.6 \dot M_{-7}^{1/2}{\rm
g\,cm^{-2}}$ if the minimal field (\ref{eq:windfield}) is used for
$B$ in eq.~(\ref{eq:betas}).

\subsection{Magnetorotational turbulence}\label{subsec:mri}

Accretion may be driven by radial transport of angular momentum
within the disk or active layer rather than the torques exerted by a
wind.  We assume that the transport is due to turbulence excited by
the magnetorotational instability (hereafter MRI). In a statistical
steady state, the constancy of the angular momentum of the disk
within radius $r$ requires
\begin{equation}
  \label{eq:Jcons}
  \dot M\Omega r^2 +2\pi r^2\int\limits_{-\infty}^{\infty}
\left(\rho\overline{v'_r v'_\phi}-\overline{B_rB_\phi}/4\pi\right)\,dz
= \Gamma_0\,.
\end{equation}
Here $\Gamma_0$ is the torque exerted on the inner edge of the disk;
at radii $\gg 0.1\au$, it is probably safe to neglect this torque
compared to the larger terms on the left hand side above. The two
terms in the integrand are the relevant components of the Reynolds
and Maxwell stresses for angular-momentum transport, with $\boldsymbol{v}'$
being the mass-weighted velocity fluctuation associated with the
turbulence, $\boldsymbol{v}'\equiv\boldsymbol{v}-\overline{\boldsymbol{v}}$ so that
$\overline{\boldsymbol{v}'}=0$.  Linear analysis and nonlinear simulations
indicate that the Reynolds stress in MRI turbulence has the same
sign as the Maxwell stress but is smaller by a factor $\sim 4$
\citep[and references therein]{Pessah_Chan_Psaltis06}, so for the
purposes of the estimates below, the Reynolds stress will be
neglected, and the Maxwell stress will be assumed to be confined to
active layers each of column density $\Sigma_a$ and thickness $h_a$
on either side of the disk midplane.

With these approximations, the average Maxwell stress within the
active layers becomes
\begin{equation}
  \label{eq:maxwell}
  -\overline{B_rB_\phi} \approx \dot M\Omega/h_a\,.
\end{equation}
Since $\overline{B_r^2+B_\phi^2}\ge2|\overline{B_rB_\phi}|$, a lower
bound for the root-mean-square total magnetic field in the active
layers is
\begin{equation}
  \label{eq:Bminmri}
  B\gtrsim 3.2\dot M_{-7}^{1/2} r_{\au}^{-11/8}\,{\rm G}\,,
\end{equation}
in which we have taken the thickness of the layer to be $h_a\approx
0.5 c_s/\Omega$ as before. At this field strength, the magnetic
pressure within the layer is comparable to the gas pressure:
\begin{equation}
  \label{eq:beta}
\frac{P_{\rm gas}}{P_{\rm mag}}=
\frac{8\pi \rho_a c_s^2}{B^2}\lesssim 1.0\Sigma_1\dot M_{-7}^{-1}
\,r_{\au}\,.
\end{equation}
The ratio (\ref{eq:beta}) is often denoted by $\beta$ in the MRI
literature but is not to be confused with the Hall parameters
(\S\ref{subsec:wind}).  The linear analysis of Wardle (1999) and the
nonlinear simulations of \citet{Sano_Stone02} indicate that MRI may
grow in stronger fields and produce stronger turbulence with the
Hall terms than without them, at least for the favorable sign of
$\boldsymbol{B\cdot\Omega}$. However, it is generally presumed that
neither linear MRI nor MRI-driven turbulence can operate unless
$P_{\rm gas}>P_{\rm mag}$ (e.g., \citealp{Kim_Ostriker00}), so
eq.~(\ref{eq:beta}) suggests that $\Sigma_a> 10\dot
M_{-7}\,r_{\au}^{-1}\,{\rm g\,cm^{-2}}$ if MRI dominates the
accretion process.  The ratio of the minimum fields
(\ref{eq:Bminmri}) and (\ref{eq:windfield}) is $\approx
\sqrt{r/h_a}$, because the turbulent stress is exerted over an area
proportional to the thickness of the active layer, whereas the wind
stress is exerted on the (much larger) horizontal surfaces of these
layers.

Inserting the lower limit (\ref{eq:Bminmri}) into the expressions
(\ref{eq:betas}) for the Hall parameters yields $\beta_i\gtrsim
0.61\,\Sigma_1^{-1}\dot M_{-7}^{1/2}\,r_{\au}^{-1/8}$, so the ions
may be marginally magnetized.  For $\beta_i>1$, the slippage between
the neutrals and the field lines is further enhanced by ambipolar
diffusion, which scales $\propto B^2$
\citep{Wardle_Koenigl93,Wardle99,Wardle07}. This is somewhat
complicated to discuss if the abundance of negatively charged grains
is comparable to that of free electrons. In view of the discussion
of magnetic and gas pressures above, however, we expect that
$\beta_i$ will remain less than unity---though not by much---at the
radii and accretion rates of interest to us, so that the Hall
diffusivity is still marginally dominant,
$\eta_H=c^2/4\pi\sigma_H\approx Bc/4\pi en_e$.  (A more accurate
treatment would likely lead to larger lower bounds on $x_e$.)  A
dimensionless measure of the coupling between the neutrals and the
magnetic field within the active layer is the magnetic ``Reynolds
number'' based on this Hall diffusivity, $Re_{M,H}\equiv c_s
h_a/\eta_H$; like \cite{Wardle99}, we presume that $Re_{M,H}\gtrsim
1$ is necessary for sustained turbulence.  In that case, $n_e\ge
cB\Omega/2\pi ec_s^2$, and therefore
\begin{equation}
  \label{eq:xeRMH}
  x_e\gtrsim 2.6\times 10^{-11}\dot M_{-7}^{1/2}\,\Sigma_1^{-1} r_{\au}^{-9/8}
\,{\rm cm^{-3}}.
\end{equation}

As noted above, the ratio of Ohmic to Hall conductivities is
$|\sigma_O/\sigma_H|\approx \beta_e$ when $\beta_e\gg1$ and
$\beta_i\lesssim1$. At the lower bound (\ref{eq:Bminmri}) for the
field strength, $\beta_e\approx 300\dot
M_{-7}^{1/2}\Sigma_1^{-1}r_{\au}^{1/8}$. It follows that the
magnetic Reynolds number (\ref{eq:Rem}) based on the Ohmic
diffusivity should satisfy $Re_M\gtrsim 100$ in order to support
MRI-driven accretion at rates $\gtrsim 10^{-8}{\rm
M_\odot\,yr^{-1}}$.

\citet{Sano_Stone02} and \citet{Turner_Sano_Dziourkevitch07} found
that a criterion based on the Elsasser number, $\Lambda\equiv
v_{Az}^2/(\eta_O\Omega)=1$, accurately traces the edge of the MRI
turbulence in their simulations. (But the former authors referred to
this quantity as ``magnetic Reynolds number".)  The Elsasser number
depends explicitly upon magnetic field as well as $x_e$.  Using
$\Lambda$ to define $\Sigma_a$ would impede comparison with most
previous studies of the ionization balance---in particular,
\citet{IlgnerNelson06a}---which have adopted thresholds in
$c_s^2/\eta_O\Omega$.  In fact, however, linear stability depends upon
at least two dimensionless parameters: $(Re_M,\Lambda)$, or
equivalently $(c_s^2/V_{Az}^2,\,\Lambda)$ since ${\rm
  Re}_M/\Lambda=c_s^2/V_{Az}^2$ The criterion ${\rm Re}_M>100$ for the
active layer is more conservative (i.e., allows a larger active layer)
when $c_s^2/V_{Az}^2\gtrsim100$, which is the case in almost all
published MRI simulations because it is desired that the
fastest-growing vertical wavelength be smaller than the gas scale
height $c_s/\Omega$. For example $c_s^2/\gamma V_{Az}^2\ge 400$ in
\citet{Sano_Stone02}, where $\gamma=5/3$ is the adopted adiabatic
index.  Moreover, this criterion is consistent with
\citet{Turner_Sano_Dziourkevitch07}'s result (see Fig. 6 of their
paper).  A strong---but not too strong---vertical field may permit linear
growth at $Re_M\sim 1$.  \citet{Gammie_Balbus94} found
instability up to $V_{Az}/c_s\lesssim 1.5$ in their ``quasi-global''
analysis of a vertically stratified shearing box, although their analysis
was limited to ideal MHD.  The nonlinear
development of MRI with equipartition-strength background fields has
not been well explored, but it is conceivable that radial advection of
flux somehow prefers such states.

\subsection{Summary of the constraints on field strength and ionization fraction}

For $\dot M\sim 10^{-7}{\rm M_{\odot}\,yr^{-1}}$, wind-driven and
MRI-driven accretion require similar ionization fractions, $x_e\sim
10^{-11}\dash10^{-10}{\rm cm^{-3}}$, as shown by equations
(\ref{eq:xeM}) and (\ref{eq:xeRMH}). Winds may allow weaker fields
than turbulence for the same accretion rate
[eqs.~(\ref{eq:windfield}) vs. (\ref{eq:Bminmri})] because of a
geometric advantage: wind torques are exerted on the disk surface,
which has a larger area than the disk thickness by $\sim r/h$.
However, winds have their own theoretical difficulties
\citep{Shu91,Ogilvie_Livio01}, and the field may have to be much
stronger than the minimal value (\ref{eq:windfield}). The constraint
that the field not exceed equipartition allows winds to couple to a
smaller surface density than turbulence
[eqs.~(\ref{eq:Sigma_min_wind}) \& (\ref{eq:beta})].

We emphasize that the constraints on field strength, active surface
density, and ionization fraction discussed in \S\ref{subsec:mri} are
based on accretion rates in the nonlinear regime rather than the
minimal conditions for linear MRI instability, which are more easily
satisfied.

\goodbreak
\section{Conductivity Calculation: Methods}\label{sec:conductivity}

In this section we describe our model for the disk conductivity,
including sources of ionization, chemical reaction rates, and grain
interactions.   Readers interested only in the results may wish to
skip ahead to \S\ref{sec:ionresults}.

Our criterion for the active layer is $Re_M\ge 100$, where
\begin{equation}\label{eq:Rem}
{\rm Re}_M=\frac{c_s^2}{\eta_O\Omega}\ ,
\end{equation}
is the magnetic Reynolds number, and
\begin{equation}\label{eq:mum}
\eta_O=230T^{1/2}x_e^{-1}\rm{cm}^2\rm{\ s}^{-1}
\end{equation}
is the magnetic diffusivity based on the Ohmic conductivity
$\sigma_O\approx e^2n_e/m_e\langle\sigma v\rangle_{en}$
\citep{Blaes_Balbus94}. The discussion of \S\ref{sec:magnetic}
indicates that the use of the single dimensionless parameter
(\ref{eq:Rem}) is an oversimplification---the Hall parameters
(\ref{eq:betas}), which involve the field strength, are also
important---but the criterion $Re_M\ge 100$ appears to be necessary
if not sufficient near $1\au$, at least if MRI turbulence dominates
the transport of angular momentum.

\subsection[]{Ionization Model}\label{ssec:ion}

A number of nonthermal processes may contribute to an excess of free
electrons over the abundance in LTE.  We consider X-rays from the
protostar and cosmic-ray ionization.

T Tauri X-rays can be very energetic, with luminosities
$L_X\approx10^{29}-10^{32}$erg s$^{-1}$ and photon energies ranging
from about 1 to 5 keV \citep{Casanova_etal95,Carkner_etal96}. We model the
X-ray source by two bremsstrahlung-emitting coronal rings
at radii $10R_{\bigodot}$ from the rotation axis and a similar
distance above and below the disk midplane
\citep{Krolik_Kallman83,Glassgold_etal97,Igea_Glassgold99,Fromang_etal02}.
The X-ray photons are attenuated by photoionization
(absorption) and Compton scattering. Scattering reflects some
of the X-rays photons from the disk, therefore reducing the total
ionization rate; but it also allows photons to penetrate deeper into
the disk by deflecting oblique rays towards normal incidence.
The photoionization cross section for
keV X-ray photons is $\sim 10^{-22}{\rm\,cm^2}$, decreases roughly
as $E_{\rm phot}^{-3}$, and
falls below the Thomson scattering cross section at $E_{\rm phot}\gtrsim$6-7~keV.
We take the X-ray ionization rate as a function of column density
normal to the disk from \citet{Igea_Glassgold99}, who
performed Monte-Carlo radiative transfer calculation
including scattering in the MMSN
model, assuming that metals are depleted onto grains and
segregated from the gas. For $T_X=3$keV, their results imply an
ionization rate per hydrogen molecule\footnote{We adopt this definition of
ionization rate throughout this paper, which is 2 times larger than
the ionization rate per hydrogen atom. Note that
\citet{Igea_Glassgold99} adopt the latter definition.} can be well fit by
\begin{equation}
\frac{\zeta_X^{\rm eff}}{L_{X,29}}\bigg(\frac{R}{1{\rm
AU}}\bigg)^{2.2}=
\zeta_1[e^{-(N_{H1}/N_1)^{\alpha}}+e^{-(N_{H2}/N_1)^{\alpha}}]
+\zeta_2[e^{-(N_{H1}/N_2)^{\beta}}+e^{-(N_{H2}/N_2)^{\beta}}]\ ,
\end{equation}
where $L_{X,29}\equiv L_X/10^{29}$erg s$^{-1}$, $N_{H1,2}$ is the
column density of hydrogen nucleus vertically above and below the
point of interest, $R$ is the cylindrical radius to the central
protostar, $\zeta_1=6.0\times10^{-12}$s$^{-1}$,
$N_1=1.5\times10^{21}$cm$^{-2}$, $\alpha=0.4$,
$\zeta_2=1.0\times10^{-15}$s$^{-1}$,
$N_2=7.0\times10^{23}$cm$^{-2}$, $\beta=0.65$. The first exponential
represents attenuation of X-ray photons by absorption, while the
second exponential incorporates a contribution from scattering. The
scaling with radius is slightly steeper than inverse square but less
steep than it would be without scattering. For $T_X=5$keV, we fit
their results with $\zeta_1=4.0\times10^{-12}$s$^{-1}$,
$N_1=3.0\times10^{21}$cm$^{-2}$, $\alpha=0.5$,
$\zeta_2=2.0\times10^{-15}$s$^{-1}$,
$N_2=1.0\times10^{24}$cm$^{-2}$, $\beta=0.7$. We have also compared
this fitted ionization rate to a direct calculation of X-ray
ionization rate without scattering based on equations (2)-(4) of
\citet{Fromang_etal02} (see Appendix A for supplemental information).
The latter gives a slightly larger
ionization rate at $1\au$. However, at larger radii (e.g. $10\au$),
scattering increases the ionization rate at columns
$\Sigma>1{\rm\,g\,cm^{-2}}$. Throughout this paper, unless stated
otherwise, we take $kT_X=3{\rm\, keV}$ and $L_X=5\times10^{29}\,{\rm
erg\,s^{-1}}$. The ionization rate depends rather weakly on the
X-ray temperature in \citet{Igea_Glassgold99}'s results [see their
Fig.~3], and much more sensitively with the pure-absorption
formalism of \citet{Fromang_etal02}. We do not entirely understand
these differences, but we base the calculations described below on
our fits to the results of \citet{Igea_Glassgold99}.


Interstellar cosmic rays (CR), unless shielded by the stellar wind,
provide ionization
\begin{equation}
\zeta_{\rm{CR}}^{\rm{eff}}=\zeta_0\exp(-\Sigma/96{\rm{g\ cm}}^{-2})\,
\end{equation}
per hydrogen atom with stopping grammage $\Sigma_0\approx96$g
cm$^{-2}$ \citep{Umebayashi_Nakano81}. Recent observations of the
cosmic-ray flux towards the diffuse cloud $\zeta$ Persei suggest
$\zeta_0\sim 10^{-16}{\rm\,s^{-1}}$ \citep{McCall_etal03}, an order
of magnitude larger than older values. We find that cosmic rays have
little effect on $\Sigma_a$ for $r<10\au$ unless $\zeta_0\gtrsim
10^{-15}{\rm\,s^{-1}}$, so we adopt $\zeta_0=0$ as our standard
value but also consider values up to
$\zeta_0=10^{-15}{\rm\,s^{-1}}$.

We neglect photo-ionization by ultraviolet photons because of their
small penetration depth. We also neglect the thermal ionization of
the alkali metals Na$^+$ and K$^+$ (e.g., \citealp{Fromang_etal02}),
which becomes important only above $\sim1000$K, and ionization by
radionuclides, which appears to be negligible.

The total effective ionization rate now is $\zeta^{\rm{eff}}=
\zeta_X^{\rm{eff}}+\zeta_{\rm{CR}}^{\rm{eff}}$. Hydrogen and helium
are the main targets, so we include only the four ionization
reactions listed Table~\ref{tab:CRion} in our chemical network.
Ionization of atomic hydrogen, neglected by IN06a, is included
because in the absence of grains and therefore of ${\rm H_2}$
formation, chemical equilibria involve H rather than ${\rm H_2}$.
The third reaction ensures that the ionization rate is almost
independent of the division between atomic and molecular phases (see
\S\ref{ssec:H2form}).

\begin{table}
\begin{center}
\begin{tabular}{llcl@{\quad}l}\hline\hline
$1.$ &  H$_2$ &  $\rightarrow$ &  H$_2^+$
 +   e$^-$
  &  $0.97\cdot\zeta^{\rm{eff}}$\\
 $2.$ &  H$_2$ &  $\rightarrow$ &  H$^+$
+   H    +
  e$^-$ &  $0.03\cdot\zeta^{\rm{eff}}$\\
 $3.$ &  H &  $\rightarrow$ &  H$^+$   +
  e$^-$
  &  $0.50\cdot\zeta^{\rm{eff}}$\\
 $4.$ &  He &  $\rightarrow$ &  He$^+$   +
  e$^-$
  &  $0.84\cdot\zeta^{\rm{eff}}$\\
\hline
\end{tabular}
\caption{Ionization reactions.  Final column is the ionization rate
  for the reaction shown.\label{tab:CRion}}
\end{center}
\end{table}


X-ray ionization is much stronger near the star, but the
X-ray flux falls as $r^{-2}$ and drops rapidly toward the disk midplane.
As a result, X-rays ionization dominates at modest radii ($r\lesssim
5\au$) and near the surface. Cosmic ray ionization is almost
negligible in the inner parts of the disk but dominates in the outer
disk and toward the midplane.



\subsection[]{Chemical Reactions}

We consider two alternative chemical reaction networks. The first is
a simplified network introduced by \citet{Oppenheimer_Dalgarno74}
for dense molecular clouds but also widely applied to protostellar
disks (e.g.,
\citealp{Gammie96,Glassgold_etal97,Fromang_etal02,Turner_Sano_Dziourkevitch07}).
This network is basically a two-element kinetic model involving five
species: a molecular species m; a neutral atomic gas-phase metal M;
their ionized counterparts ${\rm m^+}$ and ${\rm M^+}$, and free
electrons $e^-$. There are four reactions, as listed in Table~1 of
IN06a, whose reaction rates we adopt.

Following IN06a, we also consider a much extended kinetic model
involving nine elements (H, He, C, O, N, S, Si, Mg, and Fe) and
$174$ species as given by Table~A.1 of IN06a. We extract $2109$
reactions involving these species from the latest version of UMIST
database \citep{Woodall_etal07}. In our selection, we neglect photo
reactions because of the poor penetration of UV into the disk. We
also neglect ``collider'' reactions, which have very low rate
coefficients. We further exclude reactions involving CR protons and
CR photons in the UMIST database, since we have already included the
dominant reactions shown in Table~\ref{tab:CRion}. There are in
total 2113 gas-phase chemical reactions including 4 ionization
reactions in Table \ref{tab:CRion}. We have $147$ more reactions
than IN06a, who used an older version of the database \footnote{Note
that in the new version of UMIST database, several species names are
changed.}.

The rate coefficients in the database are functions of temperature.
Where the local disk temperature lies outside the stated range of
validity, we adopt the same procedure as IN06a: we replace $T_{\rm
  disk}$ with the upper or lower bound of the valid range, as
appropriate, before computing the rate.

\subsection{Reactions with Dust Grains}

Dust grains are efficient absorbers of free electrons. Grains also
interact with other neutral and ionized species mainly by
adsorption, desorption and charge exchange. Further, adsorbed
species hop on grain surfaces, and reactions can take place on grain
surfaces. There is currently no standard database for grain
reactions. We follow the prescriptions of IN06a involving ``mantle
chemistry" and ``grain chemistry". We assume that the maximum charge
of a grain particle is $2$, which is appropriate for very small
$x_e$.

{\it{Mantle species}} are adsorbed counterparts of gas-phase neutral
species. Reactions that involve mantle species belong to mantle
chemistry, as given by Table~3 of IN06a. A mantle species X[m] is
formed by collisions between a neutral species X or its ionized
counterpart X$^+$ with a grain particle, and is destroyed by
desorption. The mantle species defined here are independent of grain
charge.

Other grain-related reactions belong to grain chemistry, as given by
Table~4 of IN06a. These include charge-exchange reactions between
ionized species and negatively charged grains, absorption of free
electrons by grains, and grain charge-exchange reactions. Several
ionized species do not have any neutral counterpart (e.g. H$_3^+$,
H$_3$O$^+$), as needed in the charge-exchange reactions. Following
IN06a, we assume the products of these reactions to be the same
as for dissociative reactions in the gas phase, e.g. ${\rm H_3^+
+gr^- \to 3H+gr}$. If there are multiple final states, we adopt the
gas-phase branching ratio.


Below we describe our adopted grain properties and outline the
calculation of rate coefficients for grain reactions.

\subsubsection[]{Assumptions about Grains}\label{sssec:asp}

All grains are taken to be spherical with density $\rho_d=3{\rm
\,g\,cm^{-3}}$. Ideally, one should consider a size distribution of
grains. The conventional choice for interstellar grains is the MRN
size distribution \citep{Mathis_etal77}
\begin{equation}
N(a)da\propto a^{-3.5}da,\qquad a_{\rm{min}}\leq a\leq a_{\rm{max}}\
,\label{eq:MRN}
\end{equation}
where $a$ is grain radius, and the usual cutoffs are
$a_{\rm{min}}\approx0.005\micron$ and
$a_{\rm{max}}\approx0.25\micron$. In many previous works on disk
ionization (e.g., \citealp{IlgnerNelson06a,Salmeron_Wardle08}),
grains are chosen to have a single size. We allow up to two grain
sizes, $a_1$ and $a_2$, with different abundances. As standard
values, we take $a_1=0.01\micron$ and $a_2=0.1\micron$. The mass
ratio of the two grain populations should be
$f_1/f_2=(a_1/a_2)^{-0.5}$ to crudely represent the MRN distribution
(\ref{eq:MRN}), with $f=f_1+f_2=0.01$ as the standard mass fraction.
The grains are taken to be uniformly mixed, though we vary both the
sizes and the abundances of the two populations to imitate grain
growth and precipitation out of the surface layers.

Each grain population has its own mantle and grain chemistry.
Collisions between the two populations are allowed and cause charge
exchange in the same way as collisions among members of the same size
population. We assign equal branching ratios to the transfer of one or
two electrons.  In fact, we find that collisions between grains
belonging to different size populations are unimportant. We will show
that the electron abundance, and therefore the size of the active
zone, is dominated by the smallest grains.


\subsubsection[]{Grain reaction rates}\label{sssec:grainrates}


\paragraph{1. Collisions between ions/electrons and grains.}
These reactions correspond to reactions 7, 8 in Table~3 and
reactions 1-6 in Table~4 of IN06a. They recombine free electrons and
add to the Ohmic resistivity. The rate coefficients involve
collision rates and the probability that an electron stays on a
grain after colliding with it (sticking coefficient).

The collision rate is estimated following equations (3.1) and
(3.3)-(3.5) of \citet{Draine_Sutin87}, it is modulated by Coulomb
interactions and induced polarization.
The sticking coefficient $S$ is estimated as follows. For ion-grain
collisions $S$ is insensitive to temperature, so we take
$S_{X^+}=1$.
The sticking coefficient $S_e$ for electron-ion collisions is more
subtle. We adopt equations (B5), (B6), and (B13) of
\citet{Nishi_etal91}. Electrons undergo elastic and inelastic
scatterings (assumed to be isotropic) with grains until absorbed.
The electron sticking coefficient depends sensitively on temperature
(IN06a). For the purpose of calculating $S_e$, we take the grains to
be made of graphite, with atomic weight $12$, Debye temperature
$420$K, and electron binding energy $D_e=1{\rm\,eV}$, though $S_e$
is insensitive to the latter parameter.


\paragraph{2. Collisions between neutral gas-phase particles and grains.}
These correspond to reactions 1-5 of Table~3 in IN06a. These
reactions can potentially deplete gas-phase species by adsorbing them
onto grain surfaces, if the inverse of this process (desorption) is
slow. Similar to the previous case, the rate coefficients are
expressed by the product of collision rate and sticking coefficient.
The collision rate is just geometric. For incident particle energy
$E_i$, the probability of being adsorbed is approximately
$P_\epsilon=\exp(-\epsilon^2/2)$, where $\epsilon=E_i/\sqrt{D\Delta
E_s}$ \citep{Hollenbach_Salpeter70}; $D$  and $\Delta E_s$ denote
the dissociation energy and the amount of energy transferred to the
grain particle as lattice vibrations, respectively. For each neutral
gas-phase particle,  $D$ is approximated by its binding energy $E_D$
as given in Table A~2 of IN06a, and $\Delta E_s$ is approximated by
$2.0\times10^{-3}{\rm\,eV}$. The sticking coefficient is then
obtained by integrating $P_\epsilon$ over a thermal energy
distribution of $E_i$.


\paragraph{3. Collisions between grains.} These are reactions
7-10 in Table~4 of IN06a. They redistribute charge among grain
particles. We calculate the collision rates from equation (3) of
\citet{Umebayashi_Nakano90}. Note that we take
$\rho_d=3{\rm\,g\,cm^{-3}}$.  Since the abundance and thermal
velocities of grains are low, these reactions occur very slowly.


\paragraph{4. Desorption processes.} These are described by
reaction 6 in Table~3 of IN06a. Mantle species migrate among surface
sites at a characteristic thermal hopping frequency
\citep{Hasegawa_etal92}
\begin{equation}
\nu_0=\sqrt{\frac{2n_sE_D}{\pi^2m}}\ ,
\end{equation}
where $n_s\approx10^{15}$cm$^{-2}$ is the surface number density of
binding sites, and $m$ is the mass of the adsorbed species. The
desorption rate of mantle species via thermal hopping is then
\begin{equation}
k=\nu_0\exp(-E_D/kT_d)\ ,
\end{equation}
where $T_d$ is the temperature of the dust, taken to be the
same as the gas temperature.

Note that both adsorption and desorption rates have an exponential
factor. For adsorption rate, the exponential factor is
$\sim\exp{[-(T/T_0)^2]}$, where $T_0$ depends on $E_D$ and grain
size. For desorption rate, the exponential factor is
$\exp{(-E_D/T)}$. At relatively high temperature (e.g.
$T\gtrsim100$K for metals), the desorption rate is much higher than
adsorption rate (since $\nu_0$ is quite large), hence almost all
species are in the gas phase. As the temperature decreases, the
mantle abundance increases super-exponentially. This is particularly
pronounced for metals because they have large binding  energy $E_D$.
We give an example to demonstrate this effect, which will aid in
interpreting the results in \S\ref{ssec:g1}.

The binding energy of magnesium is $E_D=5300{\,\rm K}$.
Balancing adsorption and desorption processes, we obtain
\begin{equation}
\frac{n_{\rm Mg}}{n_{{\rm Mg[m]}}}\simeq\frac{1.34\times10^{15}\exp{(-5300{\rm
K}/T)}\exp{[(T/496{\rm K})^2]}}{(n_{\rm gr}/1{\rm cm}^{-3})\times(T/300{\rm
K})^{1/2}(a/1{\mu{\rm m}})}\ .\label{eq:desorb}
\end{equation}
For $n_{\rm gr}=0.1{\rm cm^{-3}}$, $a=0.1\micron$, and $T=300\K$,
the ratio (\ref{eq:desorb}) is about $4.1\times10^9$. However, at
$100\K$, the ratio becomes as small as $2.3\times10^{-6}$. The
critical temperature $T\approx150\K$, below which most metals are
adsorbed onto grains.

\subsection[]{H$_2$ Formation on Grain Surfaces}\label{ssec:H2form}

In the interstellar medium, grains catalyze many reactions. Adsorbed
atoms and molecules hop among sites until they collide, and the heat
of reaction is absorbed by grain lattices. Most importantly, $\Htwo$
is very efficiently formed on grains in dense molecular clouds,
where $T\lesssim20\K$. Whether $\Htwo$ formation on grains is
still important at the higher temperatures of protostellar disks is
unknown. \citet{Cazaux_Tielens02,Cazaux_Tielens04} proposed a model
incorporating both physisorption and chemisorption of H atoms on
grains. Physisorption is the adsorption process mentioned above,
whereas chemisorption involves much stronger ($\sim1{\rm\,eV}$)
chemical bonds. Radical species with unpaired electrons are likely
to be chemisorbed. At low temperatures, $\Htwo$ forms mainly by
interactions between a physisorbed and a chemisorbed H atom; at high
temperature, two chemisorbed H atoms are involved. The works cited
above found that $\Htwo$ formation can be efficient up to about
$500\K$, with efficiency up to about $0.2$. However,
\citet{Cazaux_Tielens02,Cazaux_Tielens04} overestimated their
$\Htwo$ formation efficiency.\footnote{Equations (1) and (2) of
\citet{Cazaux_Tielens04} omit a factor $\sqrt{(E-B_{ij})/E}$,
thereby overestimating the transition rate from physisorbed sites to
chemisorbed sites and overestimating the inverse transitions. Hence
the surface number density of chemisorbed H should be lower than
they claim. We have confirmed this with the author (Cazaux, private
communication).  Note that there are other typos in the formula.}

Our chemical evolution model shows that if $\Htwo$ formation is not
considered, almost all hydrogen will ultimately be converted into
atomic form, an unlikely state for a protostellar disk. Therefore,
we include $\Htwo$ formation on grains, but with low efficiency. In
view of the physical uncertainties, we model this process by
assuming a uniform probability $\eta$ for a pair of adsorbed
(physisorbed) hydrogen atoms to form an $\Htwo$ molecule. Then the
reaction rate for ${\rm H+H\rightarrow\Htwo}$ is
\begin{equation}
R=\frac{1}{2}n_Hv_Hn_d\sigma_d\eta S_H\ ,
\end{equation}
where $n_H$, $n_d$ are the number densities of H atoms and grains,
$v_H$ is the atomic thermal velocity, $\sigma_d=\pi a^2$ is the
geometric cross section of grains, and $S_H$ is the sticking
coefficient. For all $\eta>10^{-5}$, hydrogen remains molecular. We
take $\eta=10^{-3}$ as our standard value.

\subsection[]{Initial Conditions}

We normalize the abundance (proportional to number density) of
hydrogen atoms to unity, and all other elements and grains
proportionately, with the relative elemental abundances given in
Table~6 of IN06a. Elements in the gas phase other than hydrogen and
helium are depleted compared with solar abundance.  We vary the
metal (Mg, Fe) and grain abundances.
Note that we use $x_i$ to denote elemental abundances, and
$x[X]$ to denote the abundance of species X. All grains are
initially neutral, and all elements in their atomic form except
hydrogen.


\subsection[]{Numerical Method}

The chemical evolution equations are a set of first order ordinary
differential equations (ODEs). Due to a broad ranges of reaction
rates, these equations are very stiff. We use the {\tt stiff.c}
subroutine described in \citet{Press_etal92} as the main integrator.
This uses the Kaps-Rentrop algorithm, a 4th order implicit method.
The evolution equations conserve charge and elemental abundances,
but numerical errors violate the conservation laws. In our program,
we enforce conservation after each step by adjusting the elemental
abundances and $x_e$. The integrator efficiently evolves the system
for $10^{6-7}$ years.

We calculate the electron abundance by evolving the system until
chemical equilibrium is reached. Absent irreversible reactions,
chemical equilibrium would be unique and would obey detailed
balance. The inverses of about $70\%$ of the reactions in our
network are neglected, however, because these inverses are too slow
to be effective. For example, as three-body reactions are very slow
at the low densities of interest, reactions with more than $2$
products do not have any inverse in the reaction chain. Because
there are so many irreversible reactions, it is conceivable that the
equilibrium state is not unique. We have tested a few different
initial conditions, and the abundances of the main species converge
to the same values when integrated long enough. Some radiative
association reactions can be so slow that equilibrium is approached
only after very long time. Since protostellar disk lifetimes are
believed to be at most a few million years, we evolve the equations
for $10^{6}$ years and accept the final $x_e$ as the equilibrium
value (except in \S\ref{ssec:g0} where $10^7$ year is
used).\footnote{Even $10^6{\rm\,yr}$ is long compared to the
  time spent by most gas elements at constant $\rho$, $T$, and
  $\zeta$.}

We determine $\Sigma_a$, the column density of the active layer, as
follows. We assume that $Re_M$ increases monotonically upward.
If the midplane is active, $Re_M(z=0)\ge100$, then the whole column is
active. Otherwise, we start at a large disk height where
$Re_M>100$ and search by bisection to find the height at which
$Re_M=100$.

Our model parameters are summarized in Table~\ref{tab:parameters}
together with their standard values and ranges. For a single
population of grains, $a=0.1\micron$ and $f=0.01$ are the default
values. Not listed in the Table are some fixed parameters mentioned
in the text, such as the power-law indices of $T(r)$ and
$\Sigma(r)$.

\begin{table}
\begin{center}
\begin{tabular}{l@{$\dotfill$}lll}\hline\hline
 Parameter  &  Meaning  &  Standard Value &  Range \\\hline
$M_*$  &  Stellar Mass  &  $1M_{\bigodot}$  &  Fixed \\
 $\Sigma_0$  &  Disk Surface Density at $1$AU  & $1700$g\ cm$^{-2}$  &  Fixed \\
$T_0$  &  Disk Temperature at $1$AU  &  $280$K &  Fixed \\  Re$_M$  &
Critical Magnetic Reynolds Number  &  100  & $\geq$100 \\
$L_X$  &  ProtoStellar X-ray Luminosity  &  $0.5\times10^{30}$erg s$^{-1}$  &
 $10^{29-32}$erg s$^{-1}$ \\
$T_X$  &  ProtoStar X-ray Temperature  &  $3$keV  & $1-5$keV \\
$\zeta_0$  &  Cosmic-ray Ionization Rate  &  $0$s$^{-1}$  &
$\leq10^{-16}$s$^{-1}$ \\
$a_1$  & Radius of Small Grains  &  $0.01\mu$m  &
$>0.005\mu$m \\
$a_2$  &  Radius of Big Grains  &  $0.1\mu$m  &  $<3\mu$m \\
 $f_1$  &  Mass Fraction of Small Grain Particles &  $0.0076$  &  $0-0.01$ \\
$f_2$  &  Mass Fraction of Big Grain Particles  & $0.0024$  &  $0-0.01$ \\
 $\rho_d$  &  Mass Density of Grains  & $3$g cm$^{-3}$  &  Fixed \\
$\eta$ &  $\Htwo$ Formation Efficiency  &  $10^{-3}$ &  $<1$ \\
$x_{\rm{Mg}}$  & Abundance of Mg  &  $1.0\times10^{-8}$  & $<10^{-7}$ \\
$x_{\rm{Fe}}$  &  Abundance of Fe  &  $2.5\times10^{-9}$  &  $<10^{-7}$
\\\hline
\end{tabular}
\end{center}
\caption{Model parameters.}\label{tab:parameters}
\end{table}

\section[]{Conductivity Calculation: Results}\label{sec:ionresults}

In the following subsections, we start from grain-free models
(\S\ref{ssec:g0}), and gradually increase model complexity by adding
one (\S\ref{ssec:g1}) and then two (\S\ref{ssec:g2}) populations of
dust grains. In each subsection, we first discuss the chemistry,
evaluating the free-electron abundance as a function of density,
temperature, ionization rate, metal abundance, and grain properties
(if applicable). Then we apply these results to the MMSN disk model
to calculate the thickness of the active zone and its dependence on
model parameters such as X-ray luminosity, metal abundance and grain
properties. In both steps, we compare the simple chemical network
with the complex network.

\subsection[]{Models without Dust}\label{ssec:g0}

As a first step, we study the free electron abundance in pure
gas-phase chemical reaction networks. In general, this abundance
$x_e$ is a function of density ($\rho$), temperature ($T$),
effective ionization rate ($\zeta^{\rm{eff}}$), and metal abundance
($x_{\rm{Mg}}$ and $x_{\rm{Fe}}$). In the simple model, the metal
abundance is just the abundance of Mg. In the complex model, we use
the combination $x_M=x_{\rm{Mg}}+x_{\rm{Fe}}$ with
$x_{\rm{Mg}}=4x_{\rm{Fe}}$.


Our network runs very efficiently for pure gas-phase chemical
reactions. It takes less than 1000 years for the simple network to
reach chemical equilibrium. For the complex network, we find the
electron abundance is still subject to very slow variation after
evolving for $10^7$ years. Although $10^7$ year is somewhat longer
than the lifetime of the accretion phase of the protostellar disk,
this is balanced by our artificial choice of purely atomic initial
species. Therefore we choose $10^7$ year as the standard evolution
time in this subsection only.

\subsubsection[]{Chemistry}\label{sssec:g0chem}

\begin{figure}
    \centering
    \includegraphics[width=135mm,height=180mm]{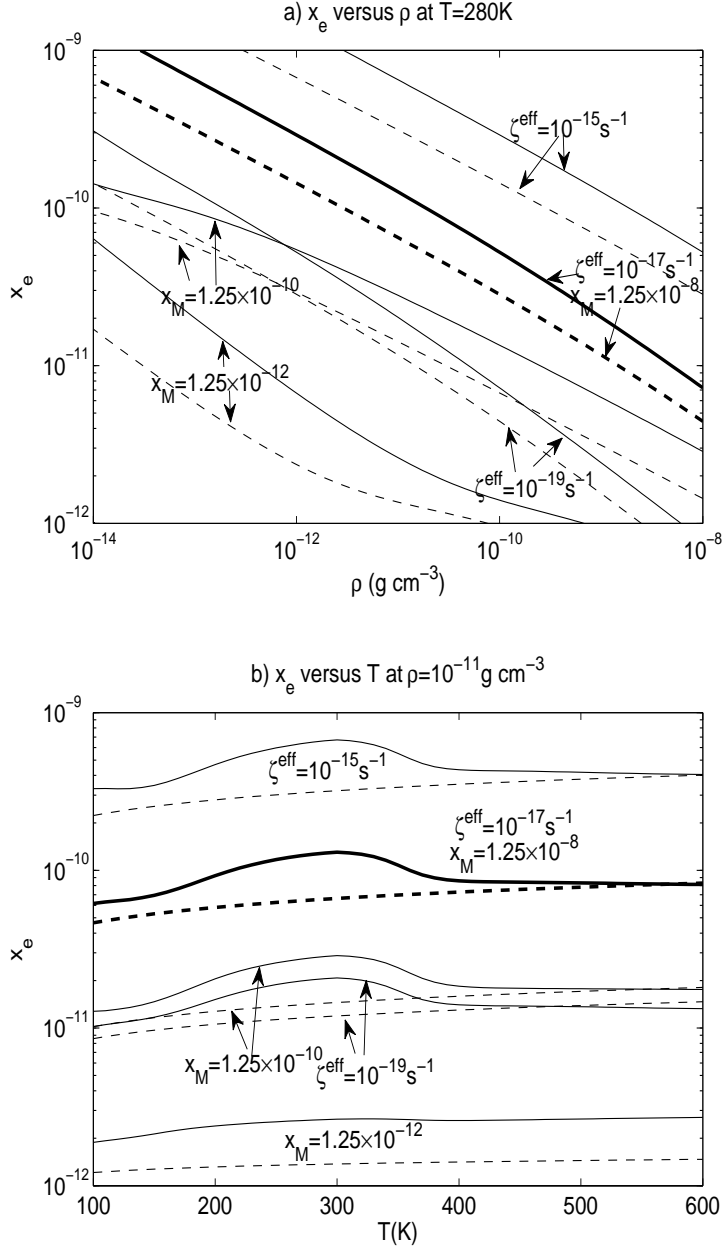}
    \caption{Electron abundance $x_e$ in grain-free models.
  (a) $x_e$ versus gas density $\rho$ at constant   $T=280\K$.
  (b) $x_e$ versus $T$ at constant $\rho=10^{-11}{\rm\, g\,cm^{-3}}$ .
    Dash lines: simple network.  Solid lines: complex network.
  Bold lines are for standard parameters: $\zeta^{\rm
    eff}=10^{-17}{\rm\,s^{-1}}$,  $x_{\rm M}=1.25\times10^{-8}$.
  Each of the other curves differs from the
  standard in one parameter, as shown. \label{fig:2}}
\end{figure}

In Figure \ref{fig:2} we plot the free electron abundance as a
function of gas density and temperature for both simple and complex
chemical models. We also vary the effective ionization rate and
metal abundance in each plot.

Note that gas densities within the disk span several orders of
magnitude. Since the (unshielded) ionization rate per unit volume is
proportional to gas density, and the (two-body) recombination rate
to density squared, one expects $x_e\propto\rho^{-1/2}$. In
Fig.~\ref{fig:2}a,  the slope is slightly flatter,
$x_e\propto\rho^{-0.4}$. By the same token, since ionization is
almost the only source of free electrons,  one expects
$x_e\propto(\zeta^{\rm eff})^{1/2}$. In both panels of Fig.
\ref{fig:2}, one sees that as $\zeta^{\rm eff}$ varies by a factor
of $100$, $x_e$ varies by $\sim7\dash10$.

One sees also that $x_e$ is insensitive to temperature. This is
because ionization reactions are independent of $T$, while recombination
reactions are exothermic, with rates roughly proportional to
$T^{-1/2}$. Most other reactions scale between $T^{-1/2}$ and $T^{1/2}$.
In protostellar disks, where $T\sim100-1000{\rm\, K}$, $T^{\pm1/2}$
varies much less than density and ionization rate and therefore
affects $x_e$ only slightly.

In these grain-free models, the electron abundance is very sensitive
to metal abundance. Metal atoms donate their electrons to ions in
charge-exchange reactions (mainly with $H^+$ and $H_3^+$); the
resulting metallic ions recombine with free electrons only by
radiative reactions, which are much slower than the dissociative
reactions available to molecular ions, e.g. ${\rm H_3^++e^->3H}$.

In most circumstances, the electron abundance obtained from the
complex network is greater than that obtained from the simple network
by roughly a factor of two. Here we differ from IN06a, who found the
opposite tendency. This is due in part to our use of a later version
of the UMIST database for the complex
network\citep{Woodall_etal07}\footnote{Also, in the absence of dust,
  the complex network gradually converts molecular to atomic hydrogen
  (see section \ref{ssec:H2form}), whereas the simple network subsumes
  hydrogen into the molecular species ${\rm m}$ and ${\rm m^+}$. Our
  addition of H ionization (see section \ref{ssec:ion}) increases the
  ionization rate. If we remove H ionization, electron abundances for
  two networks are similar.}. There are notable changes of reaction
rate in a number of reactions in the new database, and in chemical
equilibrium the abundances of several species are also quite
different.  Recently, \citet{Vasyunin_etal08} performed a sensitivity
analysis of the UMIST06 database. It was found that typical
uncertainties of molecular abundances do not exceed a factor of
3-4. The differences in $x_e$ between our results and those obtained
by IN06a with the UMIST99 database are comparable to these
uncertainties.

\begin{figure}
    \centering
      \includegraphics[width=135mm,height=180mm]{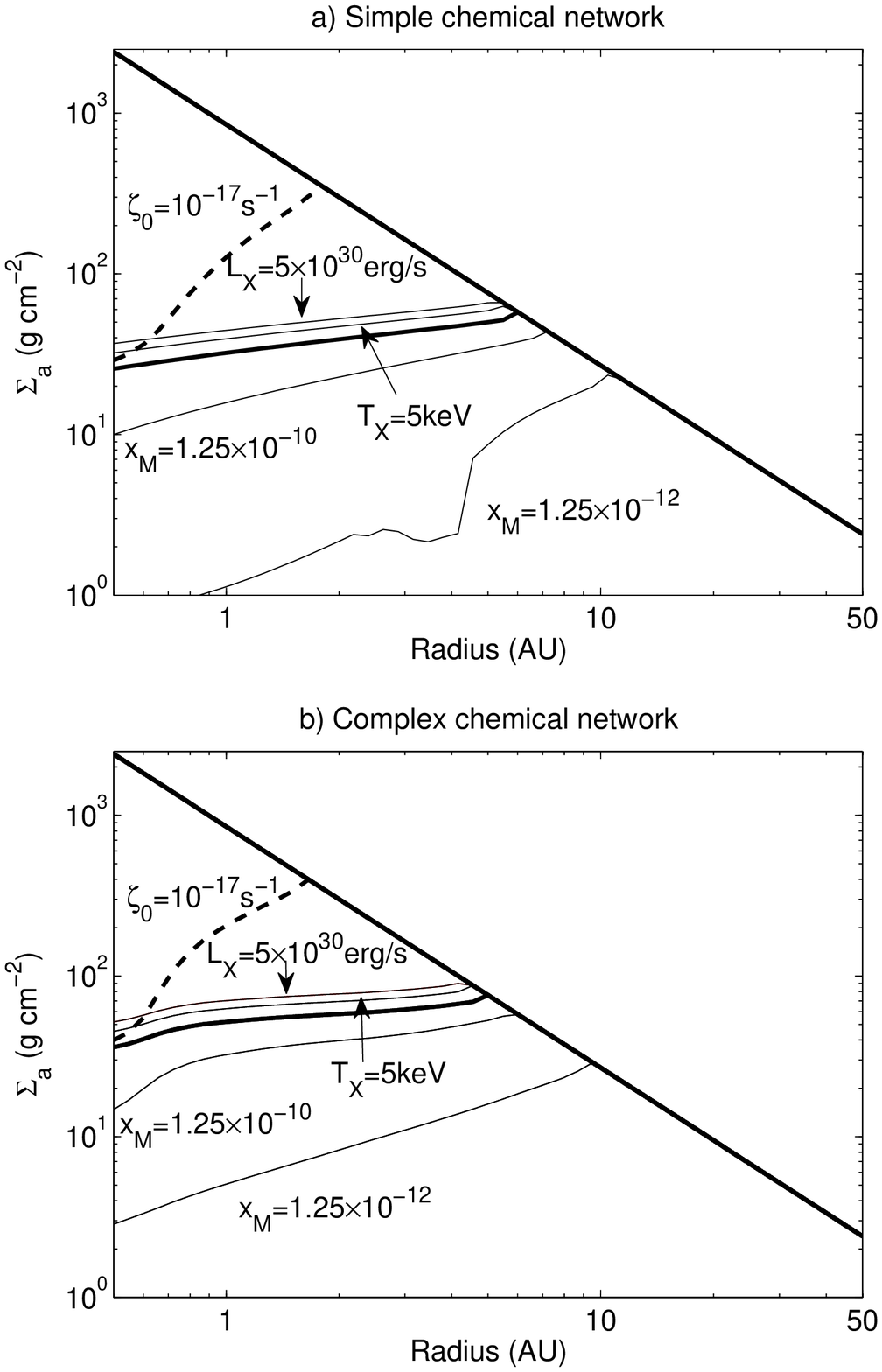}
  \caption{The column density of the active layer defined by
    $Re_M\ge100$ (\S2) versus
  radius for simple (a) and complex (b) chemical reaction networks without grain
  particles. The upper bold line indicates the half the total surface density of the disk.
 The middle bold line corresponds to our standard parameters:
  $L_X=0.5\times10^{30}{\rm\,erg\,s^{-1}}$, $T_X=3{\rm\,keV}$,
    $\zeta_0=0{\rm\,s^{-1}}$, $x_{M}=1.25\times10^{-8}$.
  Other curves differ from the standard curve in one parameter, as
  marked.\label{fig:Sig0}}
\end{figure}

\subsubsection{Application to the Disk Model}

Now temperature and density are fully fixed by the disk model, and
our free parameters are ionization rate and metal abundance. The
ionization rate is characterized by 3 parameters, namely, X-ray
luminosity ($L_X$), X-ray temperature ($T_X$) and cosmic-ray
ionization rate $\zeta_0$. Since we already know how metal
abundances and ionization rate can affect $x_e$, and we also know
how ionization parameters affect ionization rate in the disk (see
\S\ref{ssec:ion}), our main goal now is to check how these
parameters affect the column density of the active layer, $\Sigma_a$
(Note that $\Sigma_a$ is defined as the active column on one side of
the disk). In Figure \ref{fig:Sig0} we plot $\Sigma_a$ versus
radius. We also compare simple and complex chemical networks in
Fig.~\ref{fig:Sig0}.


Figure \ref{fig:Sig0} shows that in the absence of dust grains, the
requirement $\Sigma_a\ge10{\rm\,g\,cm^{-2}}$ (\S2) is easily realized.
Beyond $2\au$, the whole disk becomes active once cosmic-ray
ionization with $\zeta_0\ge10^{-17}$s$^{-1}$ is turned on. Without
cosmic-rays, X-rays can also render the whole disk active beyond
$6\au$.
In the absence of dust grains, the metal abundance strongly
influences the thickness of the active layer. At $1\au$, reduction
of $x_{\rm M}$ from $1.25\times10^{-8}$ to $1.25\times10^{-12}$
causes $\Sigma_a$ to drop by a factor $10$ for the complex network
($20$ for the simple network). Also, we see that the complex model
produces a slightly thicker active zone, consistent with the result
in \S\ref{sssec:g0chem}.

\subsection{Single-sized Grains}\label{ssec:g1}

In this subsection, we add a single population of dust grain
particles to the chemical reaction network.
The standard grain size is $a=0.1\micron$, and the standard mass
fraction is $f=0.01$. We evolve the reaction network for $10^6$ years.

\begin{figure}
    \centering
      \includegraphics[width=135mm,height=180mm]{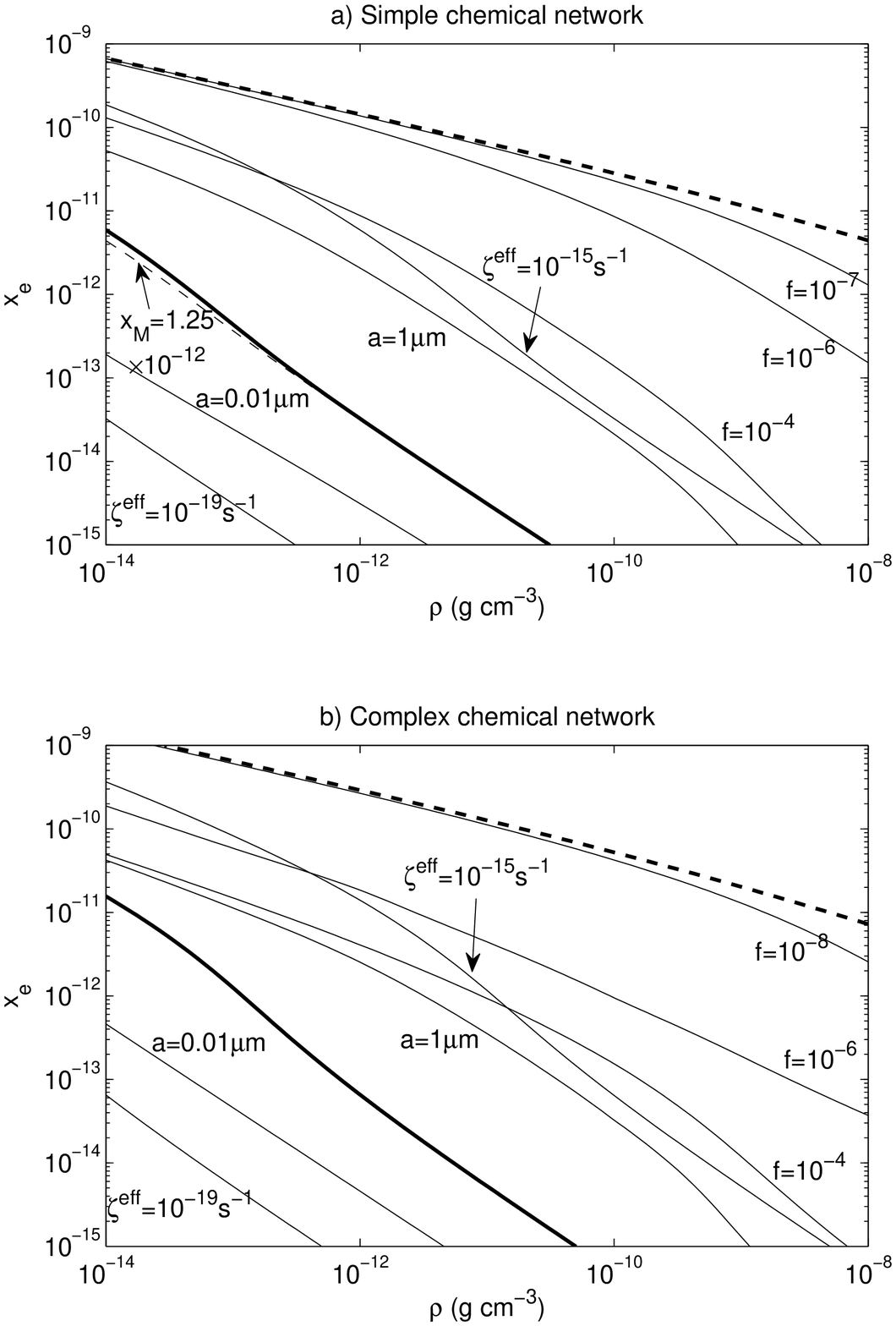}
  \caption{Electron abundance versus gas density at constant temperature,
  $T=280\K$ for simple (a) and complex (b) reaction networks with a single
  population of grains. The bold lines are for the standard values
  $\zeta^{\rm eff}=10^{-17}{\rm\, s^{-1}}$, $x_M=1.25\times10^{-8}$,
  $a=0.1\micron$, $f=0.01$, $\eta=10^{-3}$.  Other curves differ in
  one of these parameters, as marked. Bold dashed lines are grain-free
  ($f=0$). Light dashed line in (a) is for reduced metals, as shown.
  Note the tiny difference.  Panel (b) is even less sensitive to
  $x_{\rm M}$.\label{fig:xevsT1}}
\end{figure}

\begin{figure}
    \centering
    \includegraphics[width=150mm,height=105mm]{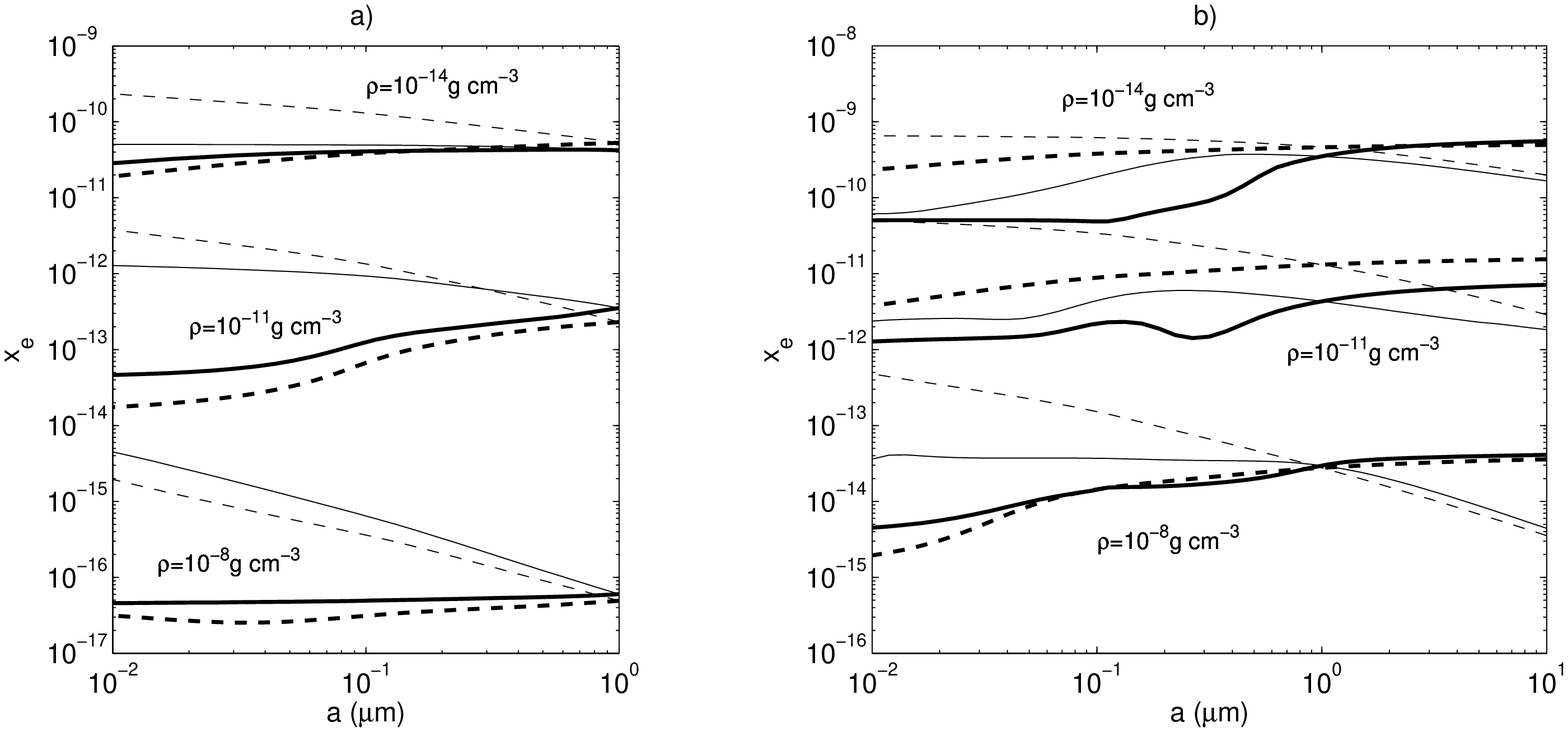}
  \caption{Electron abundance versus grain size with a single
  population of grains.  Solid lines: complex network; dashed lines:
  simple network. Plots with bold lines have fixed total grain
  area $f/a=$const, while plots with thin lines keep $f/a^2$ constant.
  Panel (a): Bold lines: $f/a=0.01\micron^{-1}$; thin lines:
  $f/a^2=0.01\micron^{-2}$. (b) Bold lines: $f/a=10^{-4}\micron^{-1}$;
  thin lines: $f/a^2=10^{-4}\micron^{-2}$. Lines in the upper, middle
  and bottom groups correspond to
  $\rho=10^{-14}$, $10^{-11}$, and $10^{-8}{\rm\,g cm^{-3}}$ respectively.
  In all plots, $T=280\K$, $\zeta^{\rm eff}=10^{-17}{\rm\,s^{-1}}$,
  $x_M=1.25\times10^{-8}$, $\eta=10^{-3}$. Note that in panel (b) the grain
  size is extended to $10\micron$\label{fig:control1}}
\end{figure}



\subsubsection[]{Chemistry}\label{sssec:g1chem}

In Figure \ref{fig:xevsT1} we plot the free electron abundance as a
function of gas density for simple and complex chemical reaction
networks respectively.
As before, we find that $x_e$ scales with density and ionization
rate, and is not very sensitive to temperature. There are also some
notable differences. Firstly, comparing the bold solid and dashed
lines, one sees that with dust grains, $x_e$ decreases faster as
density increases, suggesting that dust is more effective in
suppressing ionization in denser environments. Secondly, comparing
Fig. \ref{fig:xevsT1} with Fig. \ref{fig:2}a, one sees that changing
ionization rate by a factor of 100 causes $x_e$ to change by less
than a factor of 10 in the absence of dust, but by a factor of $25$
in its presence.

A striking difference from the grain-free case is that
gas-phase metal atoms are no longer important.
This is consistent with what IN06a have found, but our explanation
is somewhat different from theirs. At $280\K$, metal atoms are
\emph{not} swept up by grains (see \S\ref{sssec:grainrates}) at all.
Rather, the role that grains play is to promote recombination. We
find that when we add grains, the number density of metallic ions is
greatly reduced.

With our fiducial parameters, we see that the complex network
produces more free electrons, by a factor of $\lesssim2$. However,
as we gradually suppress  grains, the simple model recovers the
grain-free result more rapidly than the complex model does. In
certain ranges, the simple network can produce larger $x_e$  than
the complex one. For the simple network, reduction of grains by
$10^4$ ($f=10^{-6}$) is almost enough to recover grain-free result
in low density regions, but for complex network, depletion of nearly
$10^6$ ($f=10^{-8}$) is required.

In Figure \ref{fig:xevsT1}, we see that electron abundance increases
as grains are suppressed, and at fixed grain mass fraction, $x_e$
increases with grain size. This is as expected, but grains actually
take effect in a complicated way, especially in the complex network,
as suggested by the fact that the curves in Fig. \ref{fig:xevsT1}
are not simply vertical translates of one another. It is natural to
ask what combination of $a$ and $f$ best controls $x_e$.
To investigate this, we plot in Figure~\ref{fig:control1} $x_e$
versus grain size $a$ for a single population of grains at various
densities. Plots with bold lines have fixed total surface area
$Na^2\propto f/a=$const, while for thin lines we have fixed
$Na\propto f/a^2=$const.
If total surface area completely determines $x_e$, we would expect
that bold lines to be flat. In Fig. \ref{fig:control1}a, the bold
lines for both complex (solid) and simple (dashed) network are
approximately flat, and $x_e$ slightly decreases as $a$ decreases.
This means that total grain surface area is a good approximation to
the controlling parameter, though small grains are slightly more
effective than big ones. In Fig. \ref{fig:control1}b, as grains are
substantially depleted, we see more complicated dependencies, but
with fixed total surface area, the trend is still that $x_e$
decreases as $a$ decreases. The fact that small grains are more
efficient in reducing $x_e$ at fixed total surface area suggests
that the controlling parameter may lie somewhere between the total
surface area $Na^2\propto f/a$ and the combination $Na\propto
f/a^2$.

The results above can be qualitatively understood. The cross
sections of all grain reactions scale with grain surface area, but
they are modulated by a grain-polarization term that enhances the
cross section of smaller grains. For the reaction $e^-+{\rm
gr}\rightarrow {\rm gr}^-$, the collision rate per grain is
proportional to \citep{Draine_Sutin87}
\begin{equation}
\frac{\Gamma_{\rm col}}{\pi a^2}\propto1+\bigg(\frac{\pi
q^2}{2akT}\bigg)^{1/2} =1+0.30\bigg(\frac{300{\rm
K}}{T}\bigg)^{1/2}\bigg(\frac{1\micron}{a}\bigg)^{1/2}
\end{equation}
and for $e^-+{\rm gr}^+\rightarrow{\rm gr}$, the reaction rate is
roughly proportional to
\begin{equation}
\frac{\Gamma}{\pi a^2}\propto1+\frac{q^2}{2akT}
=1+0.056\frac{300{\rm K}}{T}\frac{1\micron}{a}
\end{equation}
Although smaller grains also have smaller sticking coefficients for
electrons, we find through numerical experiments that over a wide
temperature range, the electron sticking coefficient for an
$a=1.0\micron$ neutral grain is greater than that for an
$a=0.01\micron$ neutral grain by a factor only $\sim2$. Therefore,
the electron abundance is roughly controlled by the total surface
area of the dust grains, but smaller grains lead to smaller $x_e$ at
fixed total surface area.

Our simple analysis here considers just the direct electron
absorption by dust grains, which roughly applies to the simple
network. For the complex network, however, much more complicated
interactions between dust and all the species, as well as the
complex reactions in the gas phase make it less predictive. Our
numerical experiments above shows that in certain regimes, total
surface area controls $x_e$, but in other regimes, $na\propto f/a^2$
may be a better approximation.

One more comment on the two reaction networks. As mentioned above,
$x_e$ converges to its grain-free value more rapidly as $f\to0$ in
the simple than in the complex network. This fact is seen more
clearly by comparing the left and right panels of Fig.
\ref{fig:control1}. Before grain depletion, $x_e$ in the complex
network is typically slightly larger than that in the simple network
(left panel). As we deplete grains by a factor of 100 (right panel),
the simple network typically yields a larger $x_e$ than that of the
complex network, by up to a factor of 10.

\begin{figure}
    \centering
      \includegraphics[width=135mm,height=180mm]{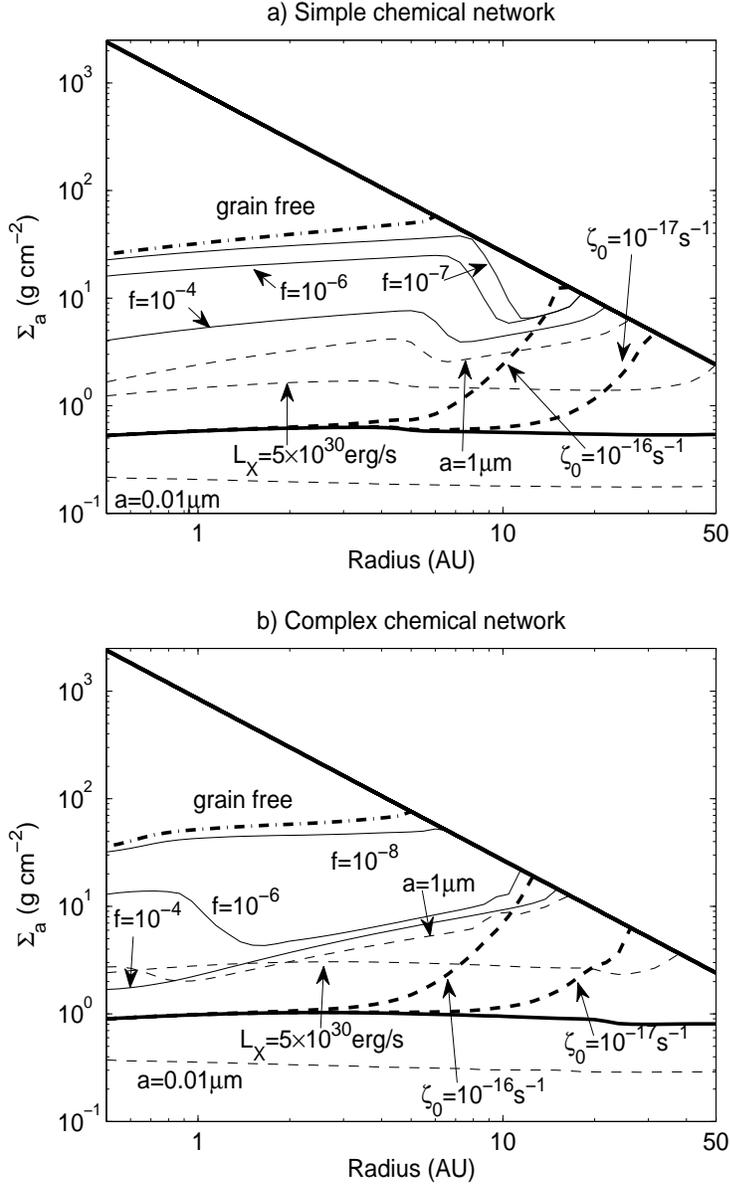}
  \caption{Like Fig.~\ref{fig:Sig0}, but for a single population of
    grains with standard values $a=0.1\mu{\rm m}$,  $f=0.01$, and
    $\eta=10^{-3}$.  All other parameters at standard values (Table~\ref{tab:parameters})
    except as indicated. For comparison, we plot the grain-free case in bold dash-dotted
    lines.\label{fig:Sig1}}
\end{figure}

\subsubsection{Application to the Disk}

Figure~\ref{fig:Sig1} plots $\Sigma_a$ against radius in the same
way as in Fig.~\ref{fig:Sig0}, and compares different model
parameters, especially grain size ($a$) and grain mass fraction
($f$). The column density of the active zone is dramatically reduced
when we add sub-micron grains. Both networks predict
$\Sigma_a\leq1{\rm\,g\,cm^{-2}}$ at $1\au$. As in Fig.
\ref{fig:xevsT1}, $\Sigma_a$ is more sensitive to ionization than in
the grain-free case. X-ray ionization alone cannot ionize the whole
disk with our fiducial parameters; but cosmic rays, if unshielded,
can ionize the entire column of the outer disk beyond $\sim10\au$.
These features are also in agreement with \citet{Sano_etal00}.

One sees that $\Sigma_a$ has a relatively simple dependence on the
grain properties in Fig.~\ref{fig:Sig1}a (i.e. simple network), but
the dependence in Fig. \ref{fig:Sig1}b is more complicated. With
fiducial parameters, $\Sigma_a$ is larger in the complex network,
but with reduced dust, the simple network can produce a thicker
active layer. Our previous conclusion that total surface area of
dust grains roughly controls the electron abundance can be checked
in the simple network, where we see that the $a=1\micron$ ($f=0.01$)
line is below the $f=10^{-4}$ ($a=0.1\micron$) line. In the complex
network, we can see that when $r<0.8\au$, the $a=1\micron$ line is
above the $f=10^{-4}$ line. This corresponds to the situation of
Fig. \ref{fig:control1}b at relatively low densities. Again, the
effect of grains on the complex network does not reduce to a single
controlling parameter.

In Fig. \ref{fig:Sig1}a we see that at about $8\au$, the two solid
lines for $f=10^{-6}$ and $f=10^{-7}$ drop sharply. This corresponds
to the sudden depletion of metals onto grains at $T\sim100\K$,
as discussed in \S \ref{sssec:grainrates}. With less dust
(smaller $f$), the transition temperature decreases [see
eq.~(\ref{eq:desorb})], and the transition radius increases. Outside
the transition radius, metals are effectively depleted onto the
grains, unless the grain number density is extremely small. In the
latter case, our grain adsorption model may fail since the surface
of all grains can be fully occupied by adsorbed species. Similar
transitions in Fig. \ref{fig:Sig1}b (e.g at $f=10^{-6}$ and
$f=10^{-8}$) are not well explained by the analysis in
\S\ref{sssec:g1chem}.

For the fiducial parameters, metals are not depleted. Under this
assumption, we see from Fig. \ref{fig:Sig1} that for $\Sigma_a$ to
be $\sim10{\rm\,g\,cm^{-3}}$ at $1\au$, $a=0.1\micron$ dust grains
must be suppressed by a factor $10^{-3}$ ($f=10^{-5}$) for the
simple network, and by $10^{-4}$ ($f=10^{-6}$) for the complex
network.  Suppression by a factor $10^{-5}$ for the simple network
and at least $10^{-6}$ for the complex network is need to recover
the grain-free case. However, if metals are already depleted (by
dust grains), we find that a depletion factor of $10^{-2}-10^{-3}$
is just enough for both the simple and complex network to recover
the grain-free result. This can also be seen in Fig.~\ref{fig:Sig1}
at larger radius where all metals are effectively adsorbed.

The differences between the networks are nontrivial. We cannot
recover the results of the complex network by rescaling parameters
in the simple one. Therefore a large array of species and reactions
may be necessary to calculating the conductivity of protostellar
disks.



\subsection[]{Two Grain Sizes}\label{ssec:g2}

There is no guarantee that the size distribution can be approximated
by grains of a single size, because different kinds of reactions
(e.g., adsorption and desorption) depend differently on grain size.
Therefore, it is necessary to explore the chemistry in protostellar
disks using more than one population of grains. On the other hand,
the behaviors of grains with different sizes are almost independent.
Neglecting grain coagulation, the direct interaction between the two
grain populations is via charge exchange, which is slow (see
\S\ref{sssec:asp}). As a result, we expect to see that
$\sum{N_ia_i^2}$ (total grain surface area) or $\sum{N_ia_i}$
roughly controls the free electron abundance, as we have found in
the previous subsection.

\begin{figure}
    \centering
    \includegraphics[width=180mm,height=135mm]{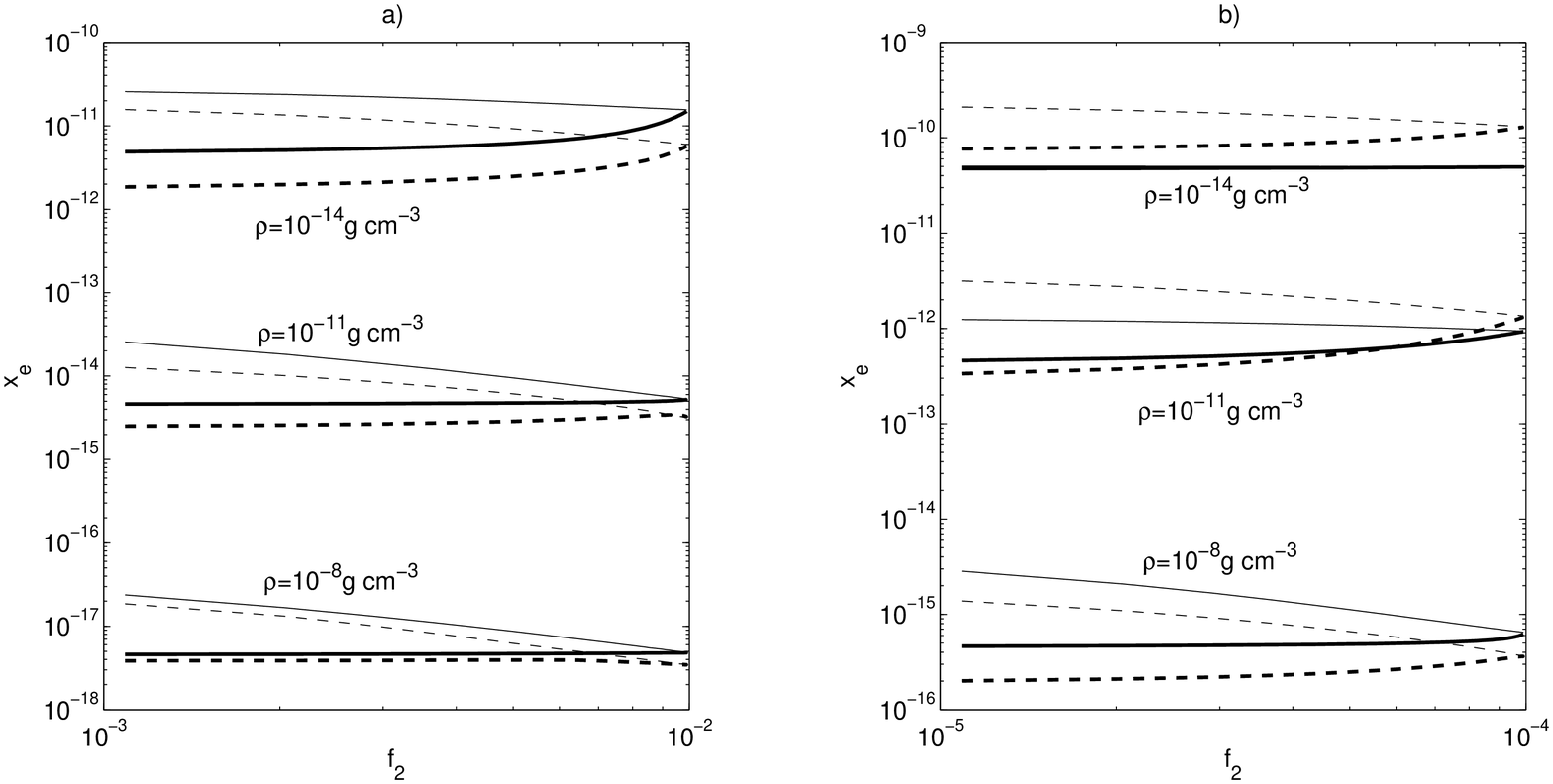}
  \caption{Like Fig.~\ref{fig:control1} but for two grain populations,
  $a_1=0.01\micron$ and $a_2=0.1\micron$.  Plots with bold
  lines have fixed total surface area $f_1/a_1+f_2/a_2=$const, while
  plots with thin lines keep $f_1/a_1^2+f_2/a_2^2$ constant. Shown
  in the plots are $x_e$ versus mass fraction of the larger grains,
  $f_2$.  (a). Bold lines: $f_1/a_1+f_2/a_2=0.1\micron^{-1}$; thin lines:
  $f_1/a_1^2+f_2/a_2^2=1.0\micron^{-2}$. (b). Bold lines:
  $f_1/a_1+f_2/a_2=10^{-3}\micron^{-1}$; thin lines:
  $f_1/a_1^2+f_2/a_2^2=0.01\micron^{-2}$.
  Note that in the upper part of panel (b), thin and
  thick solid lines almost overlap.\label{fig:7}}
\end{figure}

\begin{figure}
    \centering
    \includegraphics[width=180mm,height=135mm]{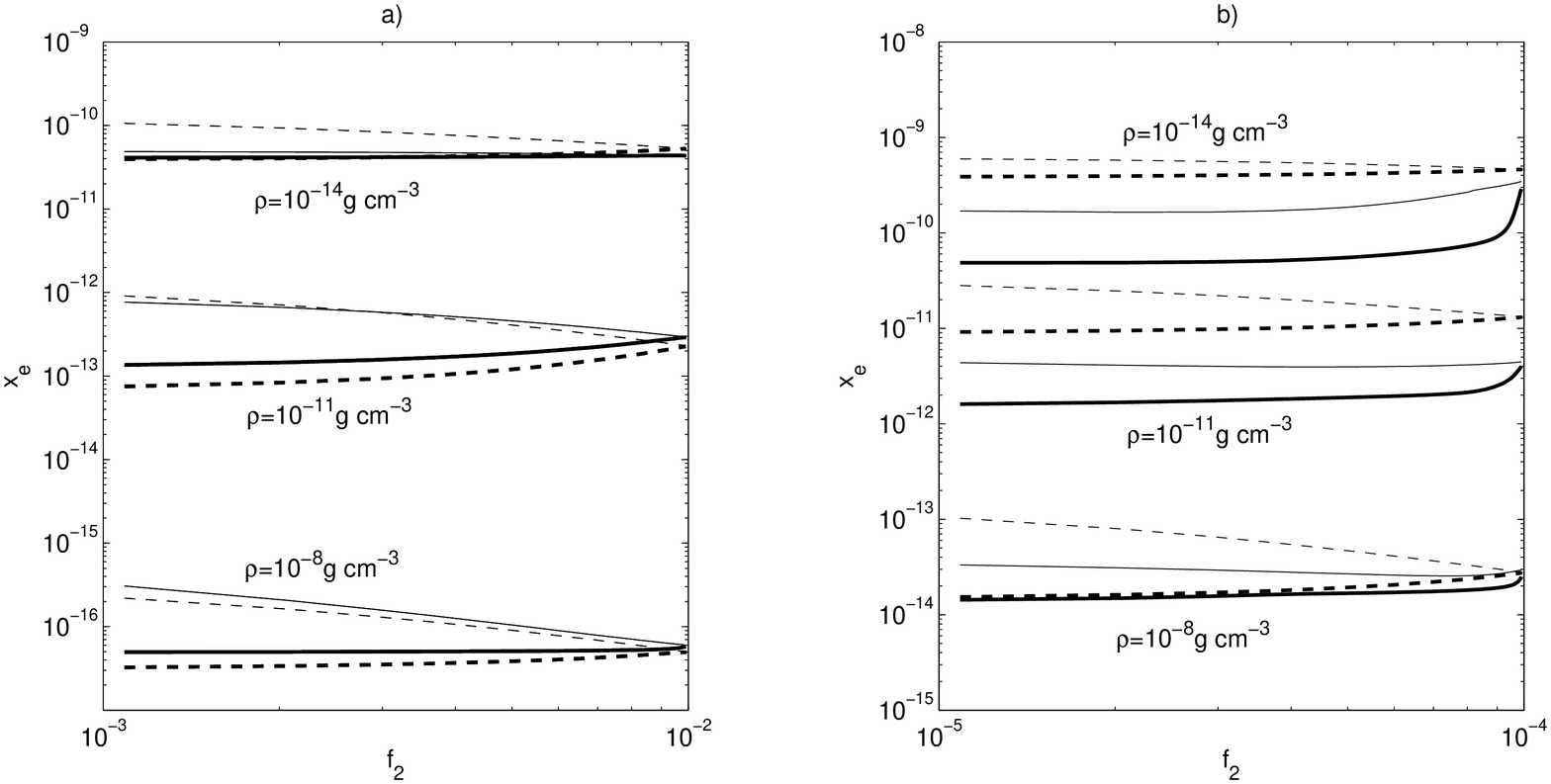}
  \caption{Same as Fig.~\ref{fig:7}, but the two grain populations are
  replaced by $a_1=0.1\micron$, $a_2=1.0\micron$. (a) Bold lines:
  $f_1/a_1+f_2/a_2=0.01\micron^{-1}$; thin lines:
  $f_1/a_1^2+f_2/a_2^2=0.01\micron^{-2}$. (b) Bold lines:
 $f_1/a_1+f_2/a_2=10^{-4}\micron^{-1}$; thin lines:
 $f_1/a_1^2+f_2/a_2^2=10^{-4}\micron^{-2}$.\label{fig:8}}
\end{figure}

\subsubsection[]{Chemistry}

Let $a_1$, $f_1$ be the size and mass fraction of the smaller
grains, $a_2$, $f_2$ be the size and mass fraction of the bigger
grains. Fixed total grain surface area means
$N_1a_1^2+N_2a_2^2\propto f_1/a_1+f_2/a_2=$const. Fixed
$N_1a_1+N_2a_2$ means $f_1/a_1^2+f_2/a_2^2=$const. In Figure
\ref{fig:7}, we consider relatively small grains, $a_1=0.01\micron$,
$a_2=0.1\micron$.
As in section \ref{sssec:g1chem}, one sees that when grains are
fully abundant, and when density is not too low, total grain surface
area controls $x_e$. In the low density region, and when grains are
substantially depleted, the controlling parameter of $x_e$ lies in
somewhere between $\sum{N_ia_i}$ and $\sum{N_ia_i^2}$ for the simple
chemical network. For the complex network, $\sum{N_ia_i}$ better
controls $x_e$ in these regimes. In Figure \ref{fig:8}, we consider
the two populations of grains to be larger, with $a_1=0.1\mu$m,
$a_2=1.0\mu$m.
We see the same trend as before. One notable difference is that for
the bold solid lines in Fig. \ref{fig:8}b (i.e., complex model at
fixed total surface area), there is a sharp increase as
$f_2\rightarrow10^{-4}$, or as $f_1\rightarrow0$, especially at low
densities. This means that adding a population of smaller grains
decreases $x_e$ substantially.

In all, the results above confirm the conclusions in
\S\ref{sssec:g1chem}. Small grains dominate the free electron
abundance, especially in the complex network.

\begin{figure}
    \centering
    \includegraphics[width=135mm,height=180mm]{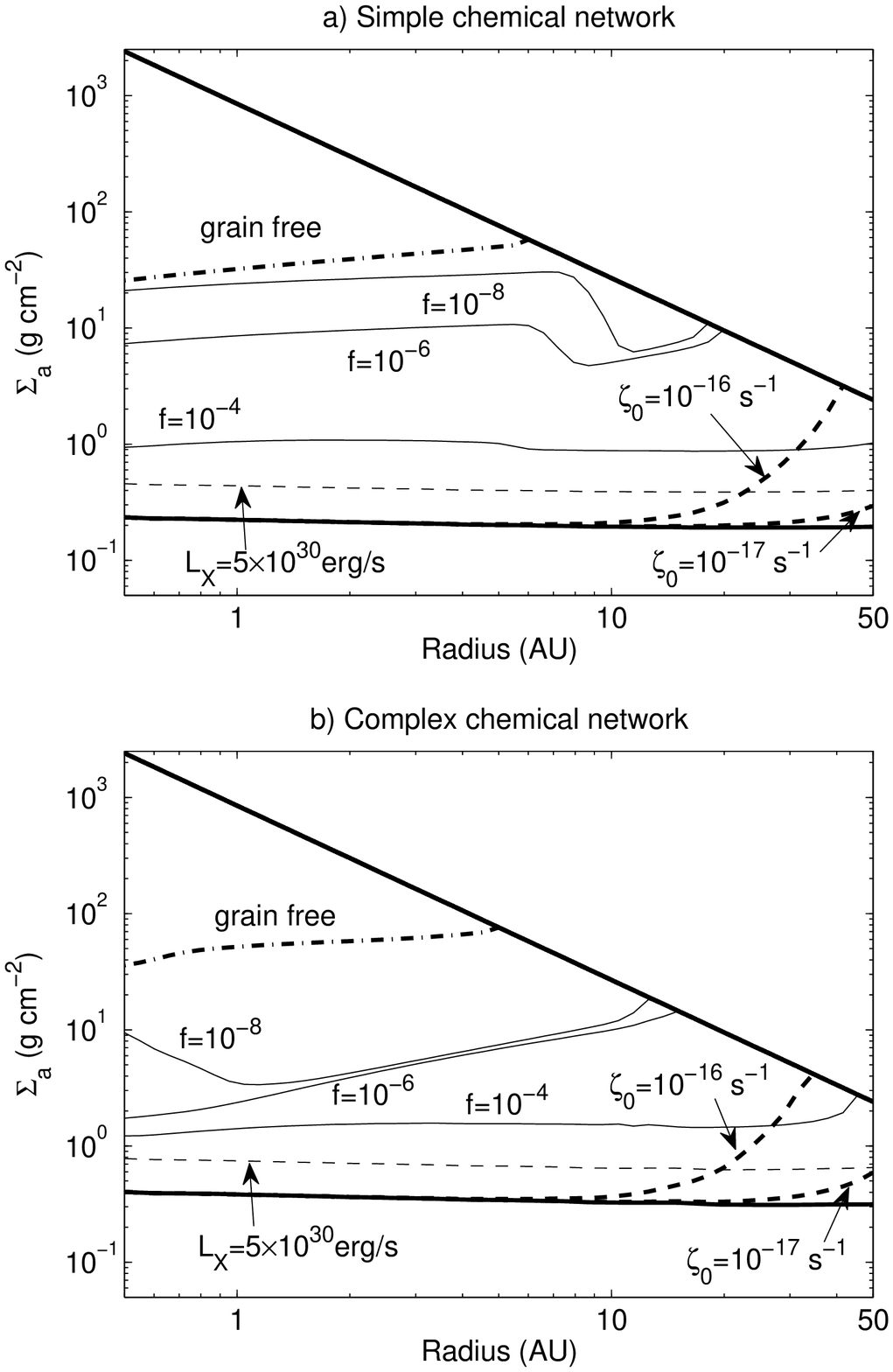}
  \caption{Like Figs.~\ref{fig:Sig0} \& \ref{fig:Sig1}, but for
  two populations of grains with sizes $a_1=0.01\micron$,
  $a_2=0.1\micron$. The mass ratio of the two populations is
  $f_1/f_2=(a_2/a_1)^{1/2}=\sqrt{10}$, in rough accord with eq.(\ref{eq:MRN}).
  We use $f=f_1+f_2$ to represent the total grain mass fraction.
  All parameters at standard values (Table~\ref{tab:parameters})
  except as indicated.\label{fig:9}}
\end{figure}

\subsubsection{Application to the Disk}


Figure \ref{fig:9} plots $\Sigma_a(r)$ for various parameter
choices.
One can see that if the dust is fully abundant ($f=10^{-2}$), the
thickness of the active layer is appreciably less than
$1{\rm\,g\,cm^{-2}}$. For the simple network, a reduction factor
$<10^{-4}$ ($f<10^{-6}$) is needed for the thickness of the active
zone to reach $10{\rm\,g\,cm^{-2}}$. For the complex network, a
reduction factor of $10^{-4}$ can raise $\Sigma_{\rm active}$ to
$3{\rm\,g\,cm^{-2}}$, whereas reduction by $10^{-6}$ ($f=10^{-8}$) is still
not sufficient to raise $\Sigma_a$ to
$10{\rm\,g\,cm^{-2}}$.
Cosmic-rays hardly affect the ionization of the disk near $1\au$.
The presence of the smallest dust grains ($a=0.01\micron$) demands a
very high ionization rate of the order $\zeta^{\rm eff}\sim10^{14}$
to achieve $\Sigma_a=10{\rm\,g\,cm^{-2}}$ for standard values of the
other parameters. This is excessively high for interstellar
cosmic-rays but might perhaps be produced by nonthermal processes
(reconnection?) within the corona of the disk or the protostar.

We conclude that when we account for the smallest grains, the column
density of the active zone is dramatically reduced. In order that the
active column $\Sigma_a$ be large enough to support vigorous accretion,
the smallest grains must be severely suppressed. In the complex
network, $x_e$ is more sensitive to the smallest grains, and more
drastic depletion is needed. Our results show that $0.01\micron$
grains should be almost completely removed, and $0.1\micron$ ones
should be reduced by a factor of $10^4$ compared to interstellar
abundances.

\goodbreak

\section{Molecular cooling of the active layer}\label{sec:cooling}

If accretion is driven by turbulent angular-momentum transport
within the disk, then in steady state, the heat released per unit
area from each side of the disk is \citep{Pringle81} $\dot Q\approx
3\dot M\Omega^2/8\pi$; the corresponding effective temperature is
\begin{equation}
  \label{eq:Tacc}
  T_{\rm eff}\approx 150\,\dot M_{-7}^{1/4} r_{\au}^{-3/4}\,\K.
\end{equation}
The heat of wind-driven accretion would be less, because of the
mechanical energy carried by the wind itself.  At a normal
interstellar gas-to-dust ratio, the heat of accretion would be
radiated primarily by the dust.  However, as we have seen, the dust
abundance must be substantially lowered in the active layer in order
that the conductivity of the layer be sufficient to couple it to the
magnetic field.  It is possible that the dust is so strongly
depleted that the layer becomes optically thin.  In that case,
molecular lines might be seen in emission from the layer.

In fact, \cite{Salyk_etal08} have observed $\water$ emission in the
$10-20\micron$ (with {\it Spitzer}-IRS) and $3-5\micron$ (with
Keck-NIRSPEC) wavelength regions in two T~Tauri systems with
particularly high accretion rates of order $10^{-6\pm 1}{\rm
  M_\sun\,yr^{-1}}$, DR~Tau and AS~205.  These authors infer that the
emitting gas lies at $\sim 1\,\au$ from the central star in both
systems and has a temperature $\sim 10^3\,{\rm K}$.  This is
probably too hot to represent the active layer as a whole unless its
optical depth is as small as $\tau<10^{-2}$ (note $T_{\rm
gas}\approx \tau^{-1/4} T_{\rm eff}$ if $\tau<1$). Indeed, their
analysis suggests a surface density $\lesssim 0.1\,{\rm g\,cm^{-2}}$
for the emitting gas, assuming a cosmic abundance of oxygen in the
form of $\water$.  So the emission may be coming from tenuous
UV-heated gas well above the active layer.  Nevertheless, these
observations further motivate us to consider the possibility of
significant molecular emission from active layers.

In order that the molecular lines should dominate the cooling,
however, the active layer must be more than modestly optically thin
to dust, that is to say, it must be that $\tau_{\rm d}\ll 1$.  The
reason is that, in contrast to the situation in planetary or stellar
atmospheres, the molecular lines are significantly narrower than
their separation, so that they cover only a small fraction of the
infrared spectrum.  In other words, the emissivity of the active
layer due to molecular lines is expected to be small. The rest of
the present section is devoted to quantifying this statement
\emph{via} a representative calculation.

Since we are not attempting to fit actual data but only to indicate
what might be expected to be emitted by a theorist's notional active
layer, we consider $\water$ only.  This is probably the most
important molecular coolant because it is expected to be abundant,
being composed of two of the most abundant elements and being
strongly thermodynamically stable under these physical conditions,
and because it has a much richer rotational spectrum than the common
linear molecules ${\rm CO}$ and ${\rm CO_2}$.  For similar reasons,
ammonia (${\rm NH_3}$) may be almost as important as water as a
coolant. Because of the terrestrial significance of water vapor, the
molecular spectrum of $\water$ is particularly well studied.  We
have relied mainly on the extensive line and energy-level lists of
\cite[hereafter BT]{Barber_etal06}, but we have also consulted the
JPL Molecular Spectroscopy database\footnote{{\tt
spec.jpl.nasa.gov}} and obtained consistent results for the
molecular emissivity at $T_{\rm gas}=300\,{\rm K}$; however, the JPL
database omits lines with wavelengths $<10\micron$, whereas $\sim
25\%$ of a blackbody emits shortward of that cutoff even at
$300\,{\rm K}$, and of course a larger fraction at higher
temperatures.

For simplicity, we represent the active layer by an isothermal slab
in LTE at gas temperature $T$ and define its emissivity by
\begin{equation}
  \label{eq:emissdef}
  \epsilon\equiv
\frac{1}{\sigma T^4}\int\limits_0^\infty
\left(1-e^{-\tau_\nu}\right)\pi B_\nu(T)\,d\nu\,,
\end{equation}
where $\tau_\nu$ is the frequency-dependent optical depth due to
water alone, which in turn is related to the water column $N $ (in
molecules~cm$^{-2}$), the line ``intensities'' $I_k$ (in
cm~molecule$^{-1}$), and the line broadening function
$\phi_k(\Delta\nu)$ (in cm) by
\begin{equation}
  \label{eq:taunudef}
  \tau_\nu= N \sum_{k} I_k \phi_k(\nu-\nu_k).
\end{equation}
Following the molecular spectroscopists' convention, the frequency
$\nu$ is measured in wavenumbers (cm$^{-1}$) rather than Hertz.  We
take $N /\Sigma_{\rm a}=3.6\times10^{20}\,{\rm molecules\,g^{-1}}$, 
which corresponds to the abundances of
\cite{Anders_Grevesse89} if all of the oxygen is in water.  The
relationship between the line intensity $I$ and the Einstein
coefficient $A_{if}$ is temperature dependent and is given for LTE
by eq.~(3) of BT.

In the limit of low columns where all lines are unsaturated
($\tau_\nu\ll 1$) eq.~(\ref{eq:emissdef}) would reduce to
\begin{equation}
  \label{eq:lowN}
  \epsilon_{\rm a}=
\frac{N }{\sigma T^4}\sum_k I_k B_\nu(T)\,,
\end{equation}
This becomes $2.17\times10^{-20}\,N$ at 300~K, corresponding to a
Planck-mean opacity $\kappa_{\rm Pl}\approx 7.8\,{\rm
cm^2\,g^{-1}}$. Since the emissivity cannot exceed unity, it is
clear that the important lines must saturate at $\Sigma_{\rm
a}>\kappa_{\rm Pl}^{-1}\approx 0.13\,{\rm g\,cm^{-2}}$, and in fact
the stronger lines saturate at even lower columns because under the
conditions of interest, the lines are very narrow.  Thus, for an
active layer of surface density $\sim 10\, {\rm g\,cm^{-1}}$, the
actual emissivity defined by eq.~(\ref{eq:emissdef}) will be
sensitive to the broadening prescription, which therefore merits
some discussion.

\begin{figure}[htbp]
  \centering
  \includegraphics[width=3.5in]{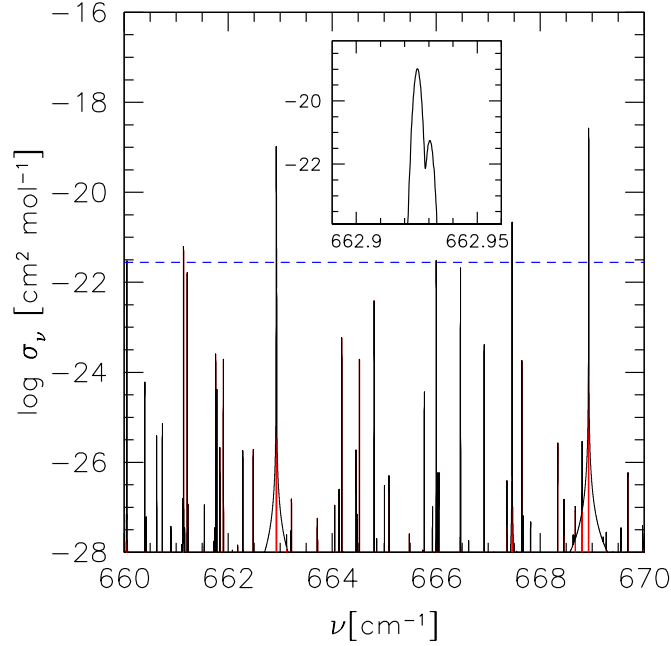}
  \caption{Frequency-dependent absorption cross-section per water molecule
    over a small range of wavenumbers $\nu=\lambda^{-1}$ at
    $T=300\,\K$ and $P=1\,{\rm \mu bar}$.  {\it Red curve:} Thermal
    Doppler broadening only.  {\it Black:} Doppler plus collisional
    broadening.  Horizontal dashed line shows level above which lines
    would be saturated for a total mass column of
    $10\,{\rm g\,cm^{-2}}$ assuming a solar abundance of oxygen in
    ${\rm H_2O}$, i.e. $3.6\times 10^{20}\,{\rm molecules \,g^{-1}}$.  Inset
    shows a closeup of two overlapping lines; but note that one is
    relatively weak.
  }
  \label{fig:H2Ospec}
\end{figure}
The relative sizes of the natural width $\Delta\nu_0$, the
collisional width $\Delta\nu_c$, the thermal Doppler width
$\Delta\nu_D$, and the turbulent width $\Delta\nu_T$ (all of these
are defined as half widths at half maximum) depend upon the
temperature, pressure, and turbulent intensity. The temperatures of
interest to us are within a factor of a few of that in the Earth's
atmosphere, $T\sim 300\times10^{\pm 0.5}\,\K$, but the pressures are
much lower. If the base of the active layer lies at a height $z_{\rm
a}\ll r$ above the midplane, then the pressure there is $P_{\rm
a}\approx \Sigma_{\rm a}\Omega^2 z_{\rm a}$; if $z_{\rm a}\approx
3[\kB T/m(H_2)]^{1/2}\Omega^{-1}$, i.e. three times the gaussian
scale height of the disk (note that $T$ here should be the
temperature near the midplane, which may be somewhat lower than in
the active layer), then
\begin{equation}
  \label{eq:Pa}
  P_{\rm a}\approx 0.7\,\left(\frac{\Sigma}{10\,{\rm g\,cm^{-2}}}\right)
\left(\frac{T}{300\,\K}\right)^{1/2}\,{\rm dyn\,cm^{-2}}\,;
\end{equation}
in other words, we are interested in pressures of order one microbar
or less $(10^6\,{\rm dyn\,cm^{-2}}=1\,{\rm bar})$. The natural width
\begin{equation}
  \label{eq:naturalwidth}
  \Delta\nu_0=\frac{1}{4\pi}\sum_{f}A_{if}\,,
\end{equation}
is $<3$~Hz (i.e., $<10^{-10}\,{\rm cm^{-1}}$ in wavenumbers) for all
upper levels $i$ tabulated by BT whose energies are less than
$(3000\,\K)\kB$ above the ground state.  This is completely
negligible compared to the other causes of line broadening; in
particular, it is much smaller than the collisional width, which
justifies our assumption of LTE. Neither BT nor the JPL database
give collisional widths.  These are difficult to calculate precisely
and are well-measured for only a minority of transitions. For our
purposes, it will be enough to have a rough estimate, so for all
lines we use
\begin{equation}
  \label{eq:nucoll}
  \Delta\nu_c= 0.075\,\left(\frac{P}{\rm bar}\right)
\left(\frac{T}{300\K}\right)^{-1/2}{\rm cm^{-1}}\,,
\end{equation}
based on the value quoted by \cite{Townes_Schawlow} for NH$_3$ (not
$\water$) in $\Htwo$.  For comparison, \cite{Giesen_etal92} find
$\Delta\nu_c=0.086\pm0.002\,{\rm cm^{-1}\,bar^{-1}}$ for $\water$ in
N$_2$ at $296\,\K$.  Thus, even at $P\sim 1\,{\rm \mu bar}$,
$\Delta\nu_c\gtrsim 10^3\Delta\nu_0$ for the important transitions.
So we have neglected the natural widths. The thermal Doppler width
of a line with central frequency $\nu_k$ is
\begin{equation}
  \label{eq:nuD}
  \Delta\nu_D = \sqrt{2\ln 2}\left(\frac{\kB T}{m(\water)\,c^2}\right)^{1/2}
\nu_k
\approx 1.46\times 10^{-3}\left(\frac{T}{300\,\K}\right)^{1/2}
\left(\frac{\nu_k}{1000\,\rm cm^{-1}}\right)\,{\rm cm^{-1}}\,.
\end{equation}
This is typically two orders of magnitude smaller than the
collisional width under terrestrial conditions, but at the much
lower pressures of an active layer, Doppler broadening should
dominate by a large factor: $\Delta\nu_D\sim 10^4\Delta\nu_c$.
Nevertheless, the collisional broadening is not entirely negligible
for us because it produces an approximately Lorentzian profile whose
wings exceed those of the thermal Maxwellian far from the line
center, and which can yield significant emission from the most
strongly saturated lines.  Finally, while the turbulent width is
uncertain because of uncertainties in the strength and nature of the
turbulence itself, for MRI-driven accretion we expect that the
turbulent Mach number should be $\lesssim 0.5[\alpha\kB T/m({\rm
H_2})]^{1/2}$.  This is less than the thermal width unless the
viscosity parameter $\alpha\gtrsim 0.5$, which seems unlikely in
these rather resistive circumstances.  On these grounds, we have
neglected the turbulent width.  But it should be borne in mind that
the turbulent velocity fluctuations might be significantly
nongaussian if the turbulence is intermittent, in which case the
wings of the turbulent velocity profile might contribute importantly
to the emission from the more saturated lines.
Figure~\ref{fig:H2Ospec} demonstrates the effects of Doppler and
collisional broadening on the frequency-dependent molecular
absorption cross section, $\sigma_\nu$; if it were due entirely to
water, the corresponding opacity would be $\kappa_\nu=\sigma_\nu N_A
X({\rm H_2O})/m({\rm H_2O})$, where $X({\rm H_2O})$ is the abundance of
water by mass, and $m({\rm H_2O})\approx 18m_p$ is the mass per
molecule. At the very low pressures (by comparison with atmospheric
or stellar conditions), relatively low temperatures, and modest
columns relevant here, the strong and saturated lines rarely
overlap. The Doppler-broadened spectrum ({\it red online}) shown in
the figure was produced with the code {\sc spectra-bt2} of BT, and
then convolved with a Lorentzian to simulate collisional broadening.

\begin{figure}[htbp]
  \centering
  \includegraphics[width=3.5in]{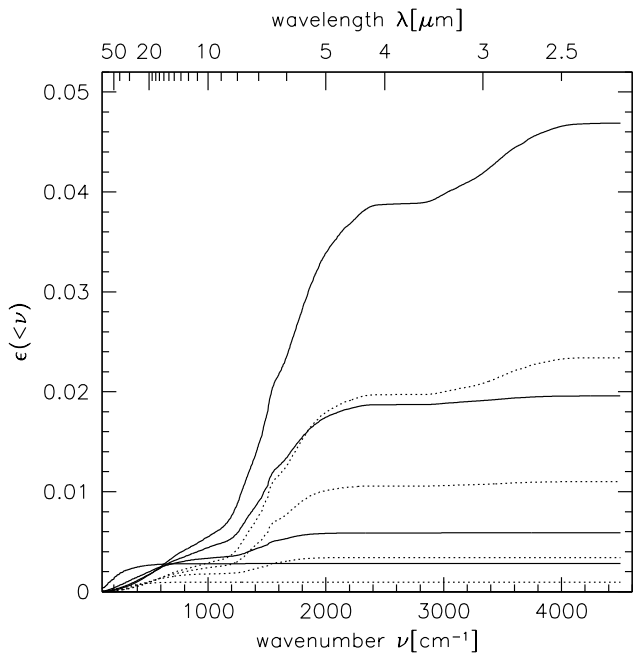}
  \caption{Cumulative emissivity of dust-free isothermal slabs due to water.
{\it Solid curves:} $\Sigma=10\,{\rm g\,cm^{-2}}$; {\it dotted:}
$1\,{\rm g\,cm^{-2}}$.
{\it Top to bottom at right:} $T=600,450,300,150\K$.
}
  \label{fig:slabs}
\end{figure}

Figure~\ref{fig:slabs} shows the cumulative emissivity defined by
the incomplete integral corresponding to eq.~(\ref{eq:emissdef}),
\begin{equation}
  \label{eq:emisscum}
  \epsilon(<\nu)\equiv
\frac{\pi}{\sigma T^4}\int\limits_0^\nu
\left(1-e^{-\tau_{\nu'}}\right) B_{\nu'}(T)\,d\nu'\,,
\end{equation}
for isothermal slabs representative of notional active layers.
Evidently, the molecular emissivity of active layers is expected to
be $\ll 1$, but perhaps not negligible compared to dust if the
latter is as strongly depleted as good magnetic coupling requires.
The sublinear dependence on column density demonstrates that the
emissivity is dominated by saturated lines.

We have found that eq.~(\ref{eq:emisscum}) cannot be calculated
reliably using the frequency-dependent opacities of
\cite{Sharp_Burrows07} because these were intended for use in
stellar and planetary atmospheres where collisional broadening is
far stronger than in disks: the saturation of the stronger lines is
therefore underestimated, so that the resulting emissivities are
typically several times larger than those shown in
Fig.~\ref{fig:slabs}.  We have not tried to use the even more recent
tabulations of \cite{Freedman_Marley_Lodders08} but would expect
similar difficulties, since those opacities were computed for
minimum pressures of $300\,{\rm \mu bar}$. Rather than construct our
own tables of $\kappa_\nu$ with the required resolution (finer than
$\Delta\nu_D/\nu\sim 10^{-6}$), we have exploited the fact that the
important lines are well separated to replace
eq.~(\ref{eq:emisscum}) with a line-by-line sum that properly
accounts for the broadening and saturation of the individual lines:
\begin{equation}
  \label{eq:emiss_lbl}
\epsilon(<\nu)\approx\frac{\pi}{\sigma T^4}\,\sum\limits_{{\rm lines}~k}
B_{\nu_k}(T)\int\limits_0^{\nu}
\left\{1-\exp\left[-NI_k\phi(\nu'-\nu_k)\right]\right\}\,d\nu'\,,
\end{equation}
It is easily seen that eq.~(\ref{eq:emiss_lbl}) strictly
overestimates eq.~(\ref{eq:emisscum}) to the extent that lines
overlap.  But direct comparisons using limited wavenumber ranges
indicate that the relative error of the line-by-line sum
(\ref{eq:emiss_lbl}) is small ($\ll 10^{-2}$) at our pressures and
temperatures.

\begin{figure}
  \centering
  \includegraphics[width=4.0in]{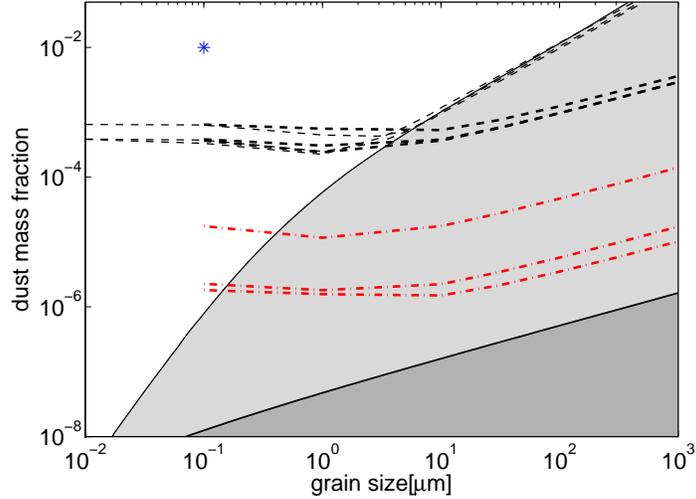}
  \caption{Constraints on grain size distribution and abundance in the MMSN at 1~AU.
    Abscissa is grain radius $a_0$ for single-size models, $N(a)\propto
    \delta(a-a_0)$, or $a_{\max}$ for MRN power-law size distribution
    (\ref{eq:MRN}) with $a_{\min}=0.01\micron$.
    {\it Light solid line:} Conductivity constraint for single-size
    grains, derived using the complex network with $\Sigma_a=10$g
    cm$^{-2}$; allowed models lie below line, {\it shaded in light grey}.
    {\it Heavy solid line:} Corresponding constraint for MRN size distribution,
    using the effective grain mass fraction of the smallest grains in equation
    (\ref{eq:constrain_pow}). Allowed region is {\it shaded in dark grey}.
    {\it Light long-dashed lines:} Loci of optical depth $\tau_{\rm dust}=1$
    for single-sized grains in active layer with $\Sigma_{\rm gas}=10\,{\rm g\,cm^{-2}}$
    and $T_{\rm gas}=150,300,600\,{\rm K}$ from top to bottom at left.
   {\it Heavy long-dashed lines:} Corresponding loci for MRN
    models.  {\it Dash-dotted lines:} Corresponding loci for equal
   frequency-averaged dust and ${\rm H_2O}$ emissivities, assuming
    MRN dust size distribution. {\it Asterisk:} Typical ISM values.}
  \label{fig:constraints}
\end{figure}

\subsection{Implication of conductivity constraints for molecular lines}

Figure \ref{fig:constraints} summarizes our constraints on the dust
in the plane of dust abundance ($f$) versus grain radius ($a$), for
single-sized grains, or maximum grain radius ($a_{\max}$) for
power-law size distributions, $n(a)da\propto a^{-3.5}da$. The
conductivity constraint is based on the complex chemical network,
with single species calculations. Parameters that lead
$\Sigma_a\geq10$g cm$^{-2}$ are considered as allowed. For the
power-law size distribution, we have not tried to reproduce the
complexities seen in Figures \ref{fig:8}-\ref{fig:9}. The
calculations show that the dependencies of the free-electron
abundance, $x_e$, and of the column density of the active layer,
$\Sigma_{\rm a}$, are not accurately given by power laws in the
dust-to-gas ratio ($f$) or grain radius ($a$). However, Figures
\ref{fig:Sig1}, \ref{fig:7}, \& \ref{fig:8} do suggest that at gas
densities relevant to the base of the active layer ($\rho\sim
10^{-11}{\rm g\,cm^{-3}}$), it is roughly the case that $x_e$ varies
with $f'(a)$ and $a$ in the combination $\int{daf'(a)/a^p}$ with an
exponent $p$ somewhere between $1$ and $2$, where $f'(a)\propto
a^3N(a)$ is the differential mass fraction of grains with size $a$.
Therefore, for the purposes of Fig.~\ref{fig:constraints}, we have
taken $\int_{a_{\min}}^{a_{\max}}{daf'(a)/a^p}$ as the controlling
parameter for the conductivity, with $p=3/2$. To account for the
power-law models with the MRN size distribution, we use a
single-size grain model at the minimum grain size $a_{\min}$ with an
effective mass fraction
\begin{equation}
f(a_{\min})=\frac{(a_{\min})^{3/2}\int_{a_{\min}}^{a_{\max}}a^{-2}da}
{\int_{a_{\min}}^{a_{\max}}a^{-1/2}da}f
=\frac{a_{\min}+\sqrt{a_{\min}a_{\max}}}{2a_{\max}}f\
.\label{eq:constrain_pow}
\end{equation}

The minimum radius is fixed at $a_{\min}=0.01\micron$, which we
believe to be conservative: if grains as small as this do exist in
the active layer, then even smaller ones are probably
present---perhaps all the way down to the PAH regime $a\sim
10\mbox{\AA}$---and such very small grains may dominate the
conductivity. However, we have not calculated any ionization
networks for grains smaller than $0.01\micron$.  Values of
$a_{\max}$ larger than $10^3\micron=1\,{\rm mm}$ are not shown in
Fig.~\ref{fig:constraints} because our own rough estimates suggest
that larger grains would precipitate out of the active layer even in
the presence of residual turbulence in the underlying ``dead'' zone
at the level $\alpha_{\rm dz}\approx 10^{-5}$, a value that we
deduce from (grain-free) simulations
\citep{Turner_Sano_Dziourkevitch07, Oishi_etal07,Turner_Sano08}.

Also shown in Figure~\ref{fig:constraints} are the loci at which an
active layer of surface density $\Sigma_{\rm a}=10\,{\rm
g\,cm^{-2}}$ would be marginally optically thick to to dust, and the
much lower loci at which the emissivities of the layer due to ${\rm
  H_2O}$ and dust would be comparable.  The dust opacities were
derived from the thermally-averaged optical efficiency factors
computed by \cite{Draine_Lee84}; we assumed a mixture of 58\%
silicate and 42\% graphite grains by mass.  The Figure shows that
the absorption opacity depends mainly on dust mass fraction provided
$a_{\max}\lesssim3\micron$, because emission and absorption then
occur mainly in the dipole regime.  The $\tau_{\rm dust}=1$ curves
are also rather insensitive to the temperature, at least over the
range we consider ($150-600\,{\rm K}$).  The equal-emissivity curves
are more sensitive, however, because of the temperature dependence
of the molecular emissivity seen in Fig.~\ref{fig:slabs}.

The main conclusion that we draw from Figure~\ref{fig:constraints}
is that the active layer should be optically thin to dust. Therefore
molecular emission lines should stand out above the dust continuum
if the layer is heated by dissipation of MRI turbulence. The
detectability of the lines will be reduced by the Doppler broadening
associated with the orbital motion of the gas, which is much larger
than thermal and collisional broadening, unless the disk is observed
face on or spatially resolved (e.g. with an interferometer).  The
upper solid line shows that if grains smaller than $a\sim 1\micron$
are entirely absent, then there does exist a regime in which
$\tau_{\rm dust}>1$ despite the ionization being sufficient for good
coupling. The lighter solid and dashed lines show that if grains smaller
than $a\sim 1\micron$ are entirely absent, then there does exist a regime
in which $\tau_{dust}\approx 1$ despite the ionization being sufficient
for good coupling. In the more plausible situation described by the 
power-law models, where small grains occur but in reduced abundance
compared to ISM values, due both to growth of the largest grains and to
reduction (or precipitation) of the overall dust mass fraction, the
conductivity constraint requires $\tau_{\rm dust}\ll 1$.  It seems even
to require that cooling of the active layer is dominated by molecules
rather than dust, unless grains smaller than $\sim0.1\micron$ are
essentially completely excluded.

\section{Discussion}\label{sec:discussion}

The preceding sections have shown that our understanding of the
interplay among grains, magnetic fields, and thermal emissions in
protostellar disks suffers from many uncertainties. Some of these
uncertainties might be reduced in the foreseeable future by further
research.  Others seem likely to plague astronomers for quite some
time.

\subsection[]{Uncertainties in the Conductivity
Calculation}\label{subsec:uncertainty}


Our criterion of the active zone is based on the choice of the
critical magnetic Reynolds number ${\rm Re}_M^{\rm crit}=100$. As
discussed in \S\ref{sec:magnetic}, our estimate may be conservative,
and our calculation may provide the upper limit of the size of the
active zone.

We have adopted the minimum-mass solar nebular disk model. This model
is idealized in two aspects. It has an empirical surface density
profile designed to match the mass distribution in the solar system.
In the MMSN, the mass density scales linearly with the surface density
coefficient $\Sigma_0$, but the disk scale height is independent of
$\Sigma_0$, and therefore so is the flux of X-rays and cosmic-rays
impinging on the disk. In fact, we have experimented with different
$\Sigma_0$, and found that $\Siga$ is essentially independent of
$\Sigma_0$, although at higher surface density, the active layer of
the disk resides at higher altitude.  The MMSN is also assumed to be
vertically isothermal. In reality, in the presence of small grains,
the surface layers should be warmer than the midplane even in a disk
passively heated by stellar irradiation
\citep{Chiang_Goldreich97,Chiang_Goldreich99}, and warmer still if
turbulent accretion is concentrated in active layers.  Higher
temperature reduces the recombination rate and thereby increases the
electron abundance, but only weakly ($\propto T^{-1/2}$, see
\S\ref{sssec:g0chem}).  Therefore, we expect that the size of the
active zone is not affected much by the disk temperature profile.
Molecular emission, however, could be strongly affected (\S\ref{sec:cooling}).

Major uncertainties attend the ionization model. The ionization rate
is proportional to the X-ray luminosity and also depends on the
X-ray temperature of the bremsstrahlung spectrum. A stellar wind may
shield cosmic rays from the disk. In the outer disk, X-ray
ionization is inefficient when small dust grains are present, and the
ionization rate critically depends on cosmic rays. These effects are
well studied in this paper. A related uncertainty comes from metal
abundances. Metal atoms are effective electron donors. In the
absence of grains, the electron abundance is directly correlated
with metal abundance. Since our results show that grains must be
severely suppressed in order for a reasonable amount of the disk to
be active, under such conditions, metals may be important. Our
standard parameter assumes a relatively high metal abundance
($x_M=1.25\times10^{-8}$). Our results also suggest that the
abundance is unlikely to be much lower than this, since an even more
severe constraint on the dust abundance would then be required.
Moreover, in the outer disk where the disk temperature is low,
essentially all the metals are adsorbed onto dust grains. Unless
cosmic-rays can provide the ionization, the outer disk is unlikely
to be very active.

Grain properties are another major source of uncertainty, as studied
in detail in this paper. We devote the next subsection discussing the
grain size distribution.  Here, we emphasize that the simple and
complex reaction networks show different response to grain
abundance. Namely, in the complete absence of grains, or at the full
ISM dust-to-mass ratio ($f=0.01$), the complex network produces more
free electrons.  However, for intermediate dust abundances, the simple
network gives a larger $x_e$, sometimes even 10 times larger.  The
Oppenheimer \& Dalgarno model, though popular for its simplicity
(eg.\citealp{Turner_Sano08}), should therefore be used with caution
when grains are assumed.

\subsection{Grain size distribution}\label{subsec:grainsize}

The basic theme of this paper is to compare constraints on the grain
abundance from the requirement of adequate magnetic coupling to
those provided by infrared observations of thermal disk emission.
For a given total mass $f$ in grains per unit mass of gas, however,
these two constraints depend differently on the sizes of grains,
to the point that no useful comparison can be made
if the size distribution is regarded as a completely free
function. Thus, it is important to consider what constraints are
available for this distribution. To bound the discussion,
we take the distribution to be a truncated power law similar to the
standard MRN form (\ref{eq:MRN}) but possibly with different lower
and upper cutoffs ($a_{\min}$ and $a_{\max}$) and with a different
exponent $p$, i.e. $N(a)\propto a^{-p}$.

Collisional cascades involving macroscopic bodies such as asteroids
or Kuijper-Belt objects result in powerlaws with $p\approx 3.5$ when
the cohesive strength is independent of size; the slope becomes
steeper (i.e. larger $p$) if large bodies are weaker than small
ones, and shallower in the opposite case \citep[and references
therein]{OBrien_Greenberg05}.  In line with this,
\citet{Jones_etal96} have proposed that the MSN distribution results
from collisions between grains overrun by shocks. In their model,
the grains are taken to have strengths similar to those terrestrial
rocks and minerals, so that the minimum relative velocity between
grains required for fragmentation is $1\,{\rm km\,s^{-1}}$.  This is
much larger than the likely collisional velocities of submicron
grains in protostellar disks.

Many authors have considered the collisional agglomeration of grains
in protostellar disks.  In a recent study,
\citet{Dullemond_Dominik05} demonstrate that if the sticking
probability is taken to be high and fragmentation is ignored, then
almost all grains smaller than $a\sim 100\,\micron$ would disappear
in less than $10^4\,{\rm yr}$, causing the disks to become optically
thin.  As they note, this contradicts the observation that many or
most T~Tauri disks remain optically thick, as judged by their
spectral energy distributions at $10-100\,\micron$, to ages of at
least $10^6\,{\rm yr}$; the suggestion is that small grains must be
replenished by fragmentation processes.

In a recent review, \citet{Natta_etal07} discuss the observational
evidence for grain growth and grain processing in protostellar
disks. Millimeter observations indicate the presence of millimeter
or even centimeter-sized grains in many disks. On the other hand,
infrared silicate features---especially crystalline features near
$10\micron$---point to the persistence of micron or submicron-sized
grains even in the oldest disks. PAH features are also sometimes
seen. But there is no clear evolutionary trend in the grain size
distribution with disk (or rather stellar) age, which points again
to active replenishment of the small-grain population.

\subsection{Turbulent Mixing}

We have calculated the free-electron abundance and conductivity as a
function of local parameters: density, temperature, grain abundance,
and ionization rate, etc. Since X-rays and cosmic rays penetrate
only a limited column, the resulting electron abundance has a large
vertical gradient. Any turbulent diffusion would mix electrons and
ions vertically. The consequences for the width of the active layer
depend upon the ratio of the timescales for mixing and
recombination. If accretion is driven by turbulence and the
turbulent transport coefficient for mass is comparable to that for
momentum (i.e. to the turbulent viscosity), then absent dust grains,
the diffusion may be faster than recombination; as a result, the
dead zone may be reduced or even eliminated \citep{IlgnerNelson06b}.

Recently, \citet{Turner_Sano_Dziourkevitch07} performed 3D MHD
simulations of protostellar disks with vertical structure, Ohmic
resistivity, and time-dependent chemical evolution based on the
simple (Oppenheimer-Dalgarno) reaction network without grains,
allowing the reactants to be passively advected. Turbulent mixing
led to a weak coupling of the dead zone to the magnetic field.
\citet{IlgnerNelson08} found in similar simulations that the effect
of turbulent mixing critically depend on gas-phase metal abundance.
\citet{Turner_Sano08} included a uniform mass fraction of $1\micron$
dust grains. Despite rapid recombination on grain surfaces, the
radial magnetic field generated in the active layer diffused toward
the disk mid-plane, causing some accretion there and rendering it
``undead".

The simulations cited above adopted microphysical parameters very
favorable to MRI. For example,
in the absence of grains, as in \citet{Turner_Sano_Dziourkevitch07}
and \citet{IlgnerNelson08}, atomic metal ions recombine only by
radiative processes, which are slow compared to the dissociative
recombination of molecular ions. The 1$\micron$ grains adopted in
\citet{Turner_Sano08} affect the electron abundance much less than
smaller grains, as we have seen. \citet{IlgnerNelson08} chose a
rather high protostellar X-ray luminosity ($10^{31}{\rm
erg\,s^{-1}}$) and temperature ($5\,{\rm keV}$) . While all of these
simulations would be described as ``vertically stratified'' in the
parlance of MRI shearing-box work because they included vertical
gravitational accelerations, buoyancy forces were excluded by the
use of isothermal equations of state. True stratification, as
expected in real disks, may inhibit vertical mixing.

Most importantly, the relevant component of the turbulent stress,
$T^{r\phi}$, was found to be at best constant with height. Since, in
steady state accretion, the accretion rate is proportional to the
areal integral of this stress [see eq.~(\ref{eq:Jcons})], the
``undead'' zones contribute to $\dot M$ only in proportion to their
geometrical thickness, which is comparable to that of the active
layers, rather than in proportion to their mass column.

\section[]{Summary and Conclusion}

We have studied constraints on small dust grains in protostellar
disks posed by the requirement that the electrical conductivity be
sufficient to support magnetically driven accretion at observed
rates. We have considered the implications of these constraints for
the production of molecular emission lines, especially at radii
$\sim 0.1-10\au$ where accretion is likely to to be confined near
the surfaces of the disk in ``active'' layers comprising a fraction
of the total column density. We adopt the minimum-mass solar nebular
(MMSN) and assume protostellar X-rays and interstellar cosmic-rays
to be the main sources of ionization. We compare a simple chemical
reaction modeled on \citet{Oppenheimer_Dalgarno74} with a complex
reaction network extracted from the latest UMIST database
\citep{Woodall_etal07}. Reactions involving dust grains are modeled
according to the general scheme laid out by \citet[hereafter
IN06a]{IlgnerNelson06a}, with some improvements, including the use
of up to two grain sizes in order to mimic an MRN grain size
distribution. The conductivity is heavily dominated by the smaller
grains. We discuss in detail the dependence of free electron
abundance ($x_e$) on density, temperature, ionization rate,
gas-phase metal abundance, and especially on grain size and grain
abundance.

Separately, we estimate that electron fractions $x_e\gtrsim
10^{-11}\dash10^{-10}$ are required at $1\au$ to explain accretion
at $10^{-7}M_\odot{\rm yr^{-1}}$, which is typical of the more
active systems; minimum magnetic field strengths range from a few
tenths of a Gauss if accretion is driven by winds, to several gauss
if it is powered by magnetorotational turbulence. It is difficult to
be very precise about these requirements because of the complexities
of winds and turbulence in the presence of vertical stratification
and tensorial conductivities, but the numbers just given are
probably conservative. We define the active layer as the column
density $\Sigma_a$ at which $Re_{M}\ge100$, where $Re_{M}\equiv
c_s/\Omega\eta$ and $\eta$ is the Ohmic diffusivity. Smaller values
of $Re_{M}$ are probably not realistic because, at the field
strengths we estimate, Hall and ambipolar drift contribute to
slippage between the neutrals and the field. By this definition, it
is difficult to achieve $\Sigma_a\ge 10\,{\rm
  g\,cm^{-2}}$ at $1\au$ even if the small grains are strongly
suppressed.

In more detail, our conclusions regarding the conductivity are as
follows:
\begin{enumerate}
\item The free electron abundance $x_e$ depends more strongly on
  density ($\rho$) and ionization rate ($\zeta$) than on temperature.
  Even a small amount of dust suppresses atomic ions such as ${\rm Mg^+}$
  and ${\rm Fe^+}$ in the gas phase, as found by \citet{IlgnerNelson06a};
  however, we find that this is due to recombination on
  grain surfaces rather than adsoprtion, except at temperatures
  ($\lesssim100K$ for Mg).

\item In the absence of grains, the complex chemical reaction network
  predicts a slightly higher  $x_e$ than the
  simple Oppenheimer-Dalgarno model, rather than the reverse as
  IN06a found. This appears to be caused by updates to the reaction
  rates in the UMIST database. These changes are within the typical
  uncertainties of those rates \citep{Vasyunin_etal08}.


\item The electron abundance depends sensitively on grains, but this
  dependence cannot be accurately characterized by a simple
  combination of grain size and grain abundance, especially in the
  complex network. Roughly however, the
  controlling parameter lies somewhere between the total grain surface
  area, $\sum{a^2N(a)}$, and the grain abundance weighted by linear
  size, $\sum{a N(a)}$.

\item Standard cosmic-ray ionization rates
  $\zeta_{CR}=10^{-17}\dash10^{-16} {\rm s^{-1}}$ render the entire disk
  column active beyond $10\au$ even in the presence of sub-micron
  grains. X-rays contribute to $x_e$ only near the disk surface,
  within columns $\Sigma\lesssim 10 {\rm g\,cm^{-2}}$. The effect of
  X-ray scattering on the ionization rate is small at $1\au$, but
  becomes important farther out.

  \item In the inner disk around $1\au$, rather extreme parameters are
    required to achieve active layers as large as $\Sigma_a=
    10{\rm\,g\,cm^{-2}}$ if submicron grains are present. For example,
    if all grains have size $a=0.1\micron$, then the dust-to-gas ratio
    must be reduced to $f\lesssim 10^{-6}$ for standard
    X-ray and cosmic-ray parameters. If $\zeta_{CR}$ were increased to
    $10^{-15}{\rm s^{-1}}$---well above ISM values but perhaps a proxy
    for nonthermal processes within
    the disk or stellar corona---then $f\lesssim 10^{-4}$ could be
    tolerated, still two orders of magnitude below the ISM abundance
    but compatible with some models of the infrared spectral energy
    distribution \citep{D'Alessio_etal06}.  If the grains
    extend to $a\lesssim 0.01\micron$ with a standard
    $N(a)\propto a^{-3.5}$ size distribution, then
    $f<10^{-6}$ even if the maximum grain size grows to $\sim 1{\rm\,mm}$
    with our standard ionization sources.

\end{enumerate}
The last two points above lead us to suspect that either an
important source of ionization has been overlooked, or else
something other than MRI turbulence drives accretion, at least near
$1\au$. However, if we suppress these suspicions, then we conclude
that the active layer at $1\au$ should be optically thin to dust in
the Planck average. Just how thin depends upon the minimum grain
size, since the dust opacity depends mainly on total dust mass
fraction $f$ as long as $a_{\max}$ is smaller than a few microns. We
have estimated the emissivity of the active layer due to water
lines, taking into account that the important lines are strongly
saturated for $\Sigma_a\gtrsim 1\,{\rm
  g\,cm^{-2}}$ because collisional broadening is almost negligible at
relevant pressures ($\lesssim 1{\rm\mu bar}$). For plausible grain
properties, the heat dissipated by MRI turbulence may be radiated
primarily in molecular lines, and the gas temperature may be
significantly higher than the effective temperature because the total
emissivity of the active layer due to dust and molecules is
small. This offers the exciting prospect that molecular lines from
active layers may be observable, and indeed may already have been
observed by \citet{Salyk_etal08}.  However, quantitative diagnoses of
active layers using such lines will require theoretical modeling of
the coupled dynamical and thermal evolution of the MRI turbulence.

\acknowledgments We thank Bruce Draine for patiently educating us
about grains and their interactions, M. Ilgner for advice concerning
the integration of large reaction networks, and S. Cazaux for
discussions of hopping rates on grain surfaces. We also thank Mark
Wardle, and our referee, Neal Turner, for stressing the
importance of X-ray scattering on disk ionization. This work was
supported in part by NSF award PHY-0821899 ``Center for Magnetic
Self-Organization in Laboratory and Astrophysical Plasmas.''

\appendix

\section[]{Comments on X-ray Ionization}

\begin{figure}
    \centering
    \subfigure{
      \label{fig:1a}
      \includegraphics[width=120mm,height=100mm]{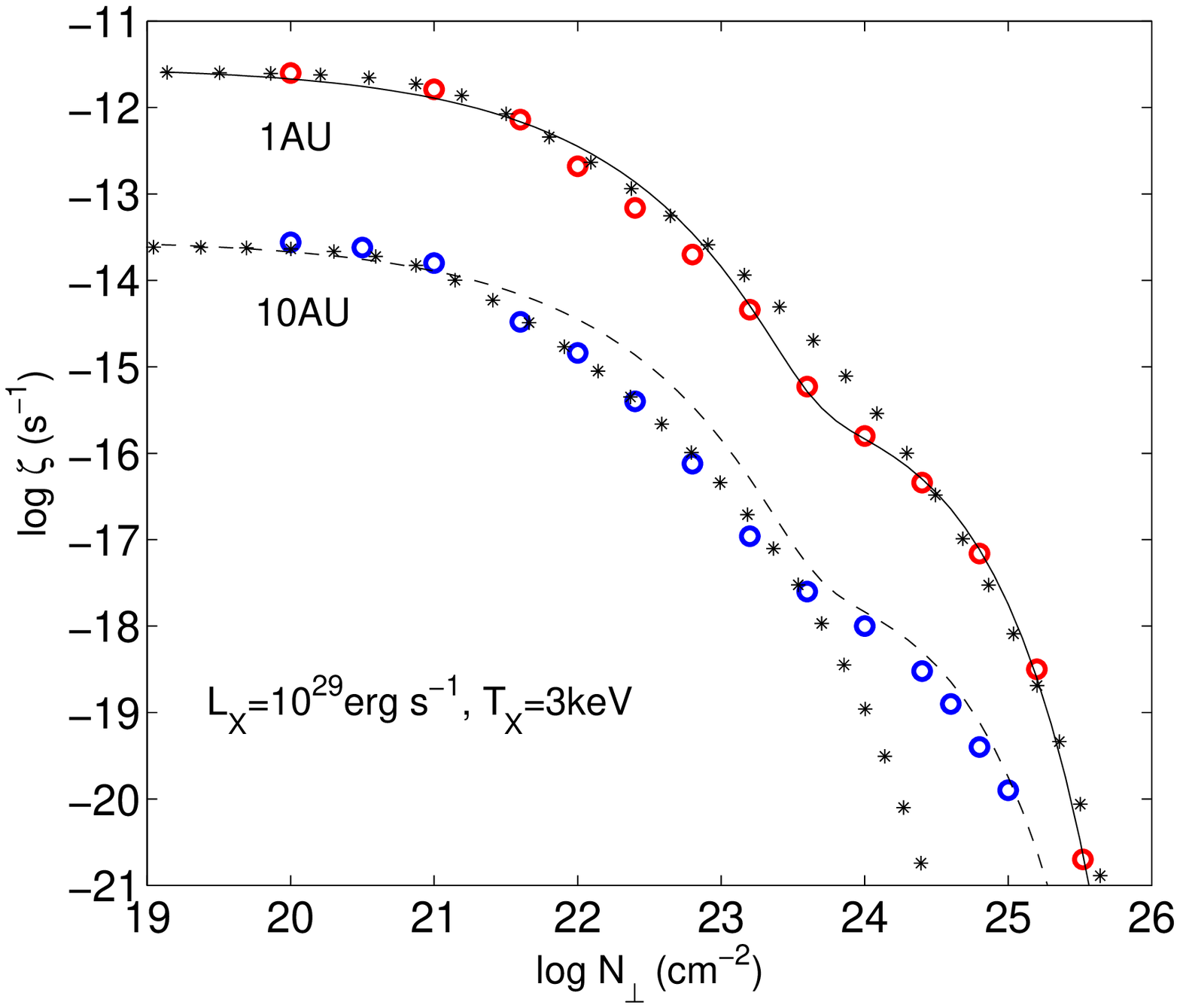}}
    \hspace{1in}
    \subfigure{
      \label{fig:1b}
      \includegraphics[width=120mm,height=100mm]{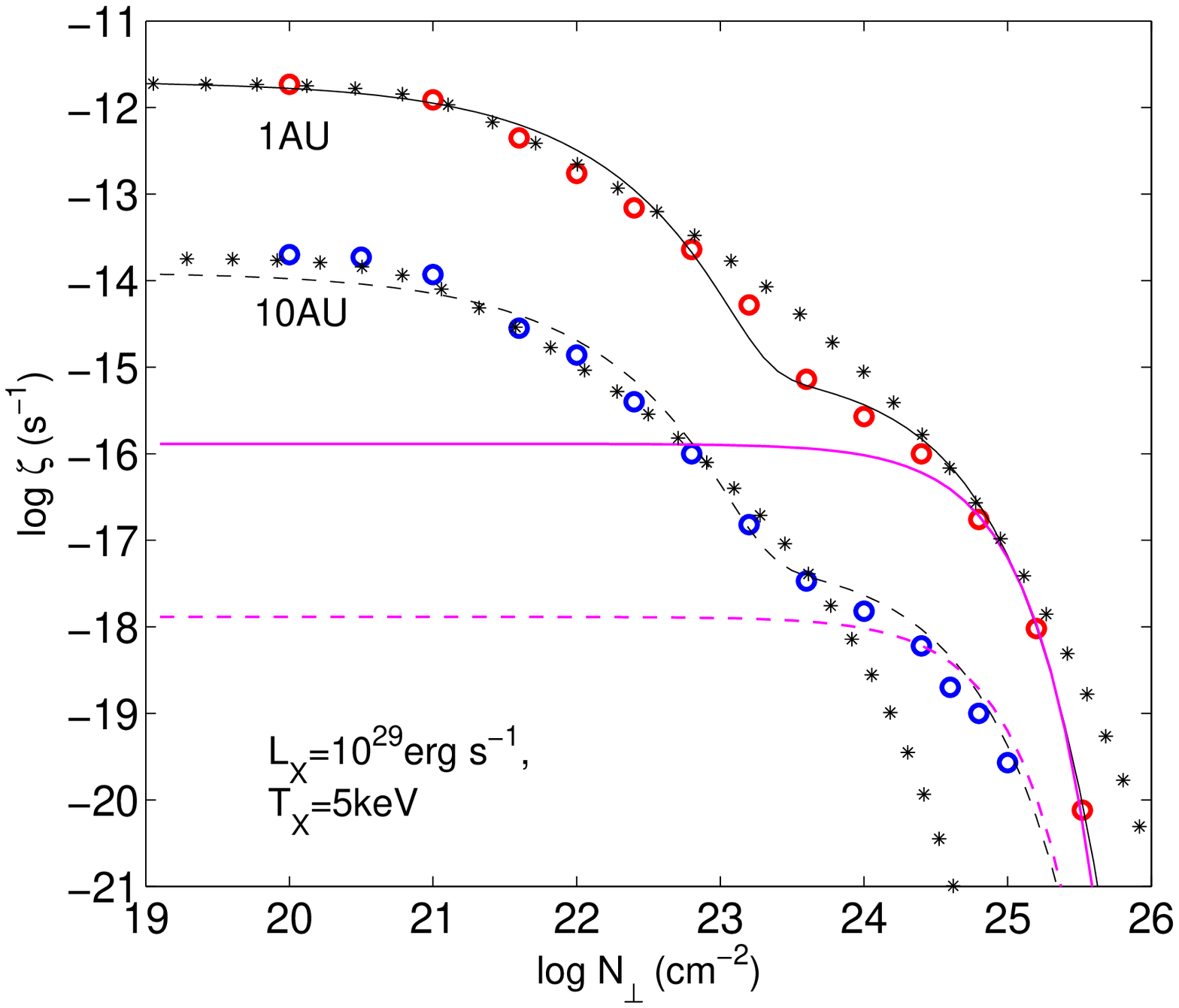}}
  \caption{X-ray ionization rate as a function of disk column density,
  at 1 AU and 10 AU respectively. The central X-ray source is assumed to
  have X-ray luminosity $L_X=10^{29}$erg s$^{-1}$, and X-ray temperature
  $T_X=3$keV (upper panel) and $T_X=5$keV (bottom panel). Circles: values taken from
  \citet{Igea_Glassgold99}. Asterisks: ionization rate with X-ray absorption
  only. Solid line: our fitted ionization rate for 1AU. Dashed line: rescaling
  of the fitted curve to 10AU. Pink curves are the fitting formula adopted by
  \citet{Turner_Sano08}. \label{fig:a1}}
\end{figure}

The X-ray ionization rate is a crucial factor on disk conductivity.
In our original version of this paper, we calculated the X-ray using
the formula given by equations (2)-(4) of \citet{Fromang_etal02}, where
the X-ray photons was assumed to be attenuated only by absorption.
However, at the energies of interest, the Compton cross section for photons
is comparable to photoionization cross section. Since X-ray photons incident
the protostellar disks obliquely, scattered photons can penetrate deeper
inward, resulting in higher ionization rate toward disk middle
plane. As supplemental information to section 3.1, we provide a comparison of
ionization rate between pure absorption model and the model including the effect
of Compton scattering \citep{Igea_Glassgold99}, as shown in Fig. \ref{fig:a1}.
This figure clarifies that neglecting Compton scattering would significantly
underestimate the ionization rate at large radii, especially towards the disk
midplane, but is OK at small radii around $1\au$.

\clearpage

\bibliographystyle{apj}

\end{document}